\newif\ifdraft \drafttrue
\newif\iffull \fulltrue

\documentclass[11pt,a4paper,final]{article}
\usepackage[left=1in,right=1in,top=1in,bottom=1in]{geometry}

\PassOptionsToPackage{numbers, compress}{natbib}

\usepackage[utf8]{inputenc} 
\usepackage[T1]{fontenc}    
\usepackage{url}            
\usepackage{booktabs}       
\usepackage{amsfonts}       
\usepackage{nicefrac}       
\usepackage{microtype}      
\usepackage{enumitem}      
\usepackage{algorithm}
\usepackage[noend]{algorithmic}
\makeatletter \@input{tex.flags} \makeatother 

\usepackage{amsmath, amssymb, amsthm}
\usepackage{mathtools}  
\usepackage{color}
\usepackage{xcolor} 
\usepackage{multicol}
\usepackage{array}
\usepackage{enumitem}   

\usepackage{flafter} 

\usepackage{tabularx}

\definecolor{OliveGreen}{rgb}{0,0.6,0}
\definecolor{lightred}{rgb}{1, 0.8, 0.8}
\definecolor{lightgreen}{rgb}{0.8, 1, 0.8}
\definecolor{blue}{rgb}{0, 0, 0}

\definecolor{DarkGreen}{rgb}{0.1,0.5,0.1}
\definecolor{DarkRed}{rgb}{0.5,0.1,0.1}
\definecolor{DarkBlue}{rgb}{0.1,0.1,0.5}
\usepackage{hyperref} 
\hypersetup{backref=true,       
    pagebackref=true,               
    hyperindex=true,                
    colorlinks=true,                
    breaklinks=true,                
    urlcolor= black,                
    linkcolor= blue,                
    bookmarks=true,                 
    bookmarksopen=false,
    filecolor=black,
    citecolor=blue,
    linkbordercolor=blue
}

\usepackage{xspace}
\usepackage{cleveref}

\usepackage{thmtools} 
\usepackage{thm-restate}
 
\usepackage{natbib}
\usepackage{tikz}
\usepackage{bbding}
\usepackage{float}
\usepackage{dsfont}

\usepackage{hyperref}       

\usetikzlibrary{positioning}

\newcommand{\xhdr}[1]{\vspace{2mm} \noindent{\bf #1}}

\usepackage[suppress]{color-edits}
 \addauthor{sw}{blue}
 \addauthor{vs}{red}
 \addauthor{ls}{DarkGreen}
 \addauthor{dn}{magenta}

\title{Incentivizing Compliance with Algorithmic Instruments}

\author{Logan Stapleton\thanks{Corresponding author: stapl158@umn.edu}\protect\phantom{\footnotesize 1}\thanks{University of Minnesota} \and Hao-Fei Cheng\thanks{Carnegie Mellon University} \and Anna Kawakami\footnotemark[3] \and Venkatesh Sivaraman\footnotemark[3] \and Yanghuidi Cheng\footnotemark[3] \and Diana Qing\thanks{University of California, Berkeley} \and Adam Perer\footnotemark[3] \and Kenneth Holstein\footnotemark[3] \and Zhiwei Steven Wu\footnotemark[3] \and Haiyi Zhu\footnotemark[3]}

\date{April 29, 2022}

\usepackage{graphicx} 
\usepackage{subcaption}

\usepackage{multirow}
\usepackage{xcolor,colortbl}
\usepackage{balance}
\usepackage{flushend}

\definecolor{OliveGreen}{rgb}{0,0.6,0}
\definecolor{lightred}{rgb}{1, 0.8, 0.8}
\definecolor{lightgreen}{rgb}{0.8, 1, 0.8}
\definecolor{blue}{rgb}{0, 0, 0}
\definecolor{tempblue}{RGB}{207,226,243}
\definecolor{tempviolet}{RGB}{234,211,220}

\begin{document}

\title{Extended Analysis of ``How Child Welfare Workers Reduce Racial Disparities in Algorithmic Decisions''}

\maketitle

\begin{abstract}
\noindent This is an extended analysis of our paper ``How Child Welfare Workers Reduce Racial Disparities in Algorithmic Decisions'' \cite{cheng2022disparities}, which looks at racial disparities in the Allegheny Family Screening Tool \cite{vaithianathan2017}, an algorithm used to help child welfare workers decide which families the Allegheny County child welfare agency (CYF) should investigate. On April 27, 2022, Allegheny County CYF sent us an updated dataset and pre-processing steps. In this extended analysis of our paper, we show the results from re-running all quantitative analyses in our paper with this new data and pre-processing. We find that our main findings in \cite{cheng2022disparities} were robust to changes in data and pre-processing. Particularly, the Allegheny Family Screening Tool on its own would have made more racially disparate decisions than workers, and workers used the tool to decrease those algorithmic disparities. Some minor results changed, including a slight increase in the screen-in rate from before to after the implementation of the AFST reported our paper \cite{cheng2022disparities}.
\end{abstract}

\section{Data \& Pre-processing Changes}
\vspace{-0.5cm}
Our paper ``How Child Welfare Workers Reduce Racial Disparities in Algorithmic Decisions'' \cite{cheng2022disparities} looks at racial disparities in the Allegheny Family Screening Tool \cite{vaithianathan2017}, an algorithm used to aid child welfare workers deciding which families the county government should investigate. The paper uses a mixed methods approach, reporting both quantitative and qualitative findings. The quantitative findings are based on real, anonymized data on children who were reported to the Allegheny County Office of Children, Youth, and Families (CYF), the AFST scores that these children received, whether the children were screened in for investigation or not, and whether the children were placed in foster care.


We retrieved the data \lsdelete{from a folder named stanford\_data} within a secure file location in Carnegie Mellon University used to store sensitive data received from Allegheny County CYF, which is maintained by colleagues of ours. The primary data we used in our paper was stored in files named PAN\_Retro\_Run\_Referrals\_for\_2014-2016\_provided\_2018-08-29.csv and PAN\_Retro\_Run\_08012016\_07132018.csv. At the time of writing our paper, we believed this data to be the same as those used in an CYF-commissioned Impact Evaluation of the AFST done by Stanford researchers \lsdelete{(hence the folder named ``stanford\_data'') }\cite{goldhaber2019impact}. As such, we preprocessed the data in the same way that \citet{goldhaber2019impact} describe in their paper: First, we excluded all referrals not marked General Protective Services (GPS) by deleting all data entries where the variable REFER\_TYPE\_GPS\_NULL was not 1. Second, we excluded referrals connected to active and closed cases by deleting entries where the variable CALL\_SCRN\_OUTCOME was either `Accept: Actively working with this family' or `Screen Out: **Assessment Completed on Active Family**' in a secondary dataset named LIM\_REFERRAL\_CLIENTS\_08262021.csv. We exclude referrals which the AFST would not have had an impact on: non-GPS referrals and referrals that are connected to active cases are automatically screened in by workers, so the AFST would have had no impact (because workers had no choice).


On April 19, 2022, we received notice from employees at Allegheny County CYF that they have been using a different dataset and pre-processing steps for their internal analysis. They recommended that we not use the PAN\_Retro files (referred to above) to define the population of referrals, the CALL\_SCRN\_OUTCOME variable to exclude active cases, nor the MCI\_ID variable we used to identify each individual child. Particularly, they said the MCI\_ID variable contained some errors, where one child could be assigned multiple children. Instead, they recommended that we use a new variable called MCI\_UNIQ\_ID, which is similar to MCI\_ID but removes many of these erroneous redundancies, to identify individual children; and that we use a new variable called ACTIVE\_FAMILY\_IND to identify active cases. Neither MCI\_UNIQ\_ID nor ACTIVE\_FAMILY\_IND were available to us at the time we wrote our paper. On April 21, 2022 we had an online video meeting with a member from the CYF Child Welfare Analytics team and received confirmation on the new pre-processing steps we would take to re-run our analysis. On April 27, 2022, a CYF employee sent us a new dataset LIM\_REFERRAL\_CLIENTS\_UNIQapp\_04212022\_v3.csv which includes information on all referrals from January 1, 2015 through July 13, 2018, including the ACTIVE\_FAMILY\_IND and MCI\_UNIQ\_ID variables. They also sent us a file called RETRO\_FILES\_COMBINED\_04212022\_v3.csv which contains retroactively-run AFST scores (which corrected for a glitch in AFST V1 scores \cite{De-Arteaga2020}) for children reported during this time period. Following CYF’s recommendations to use this new data they sent us, as well as which pre-processing steps to use, we re-ran all of the quantitative analyses in our paper. For clarity, we will refer to the data, pre-processing, and quantitative analysis in our original paper as \textbf{Analysis 1} and this second analysis based on new data and pre-processing as \textbf{Analysis 2}.\footnote{All of our code is publicly available at \url{https://github.com/logan-stapleton/AFST_racial_disparity}.}

\section{Main Takeaways}
\begin{figure*}[t!]
    \centering
    \begin{subfigure}[c]{0.5\textwidth}
        \centering
        \includegraphics[width=\linewidth]{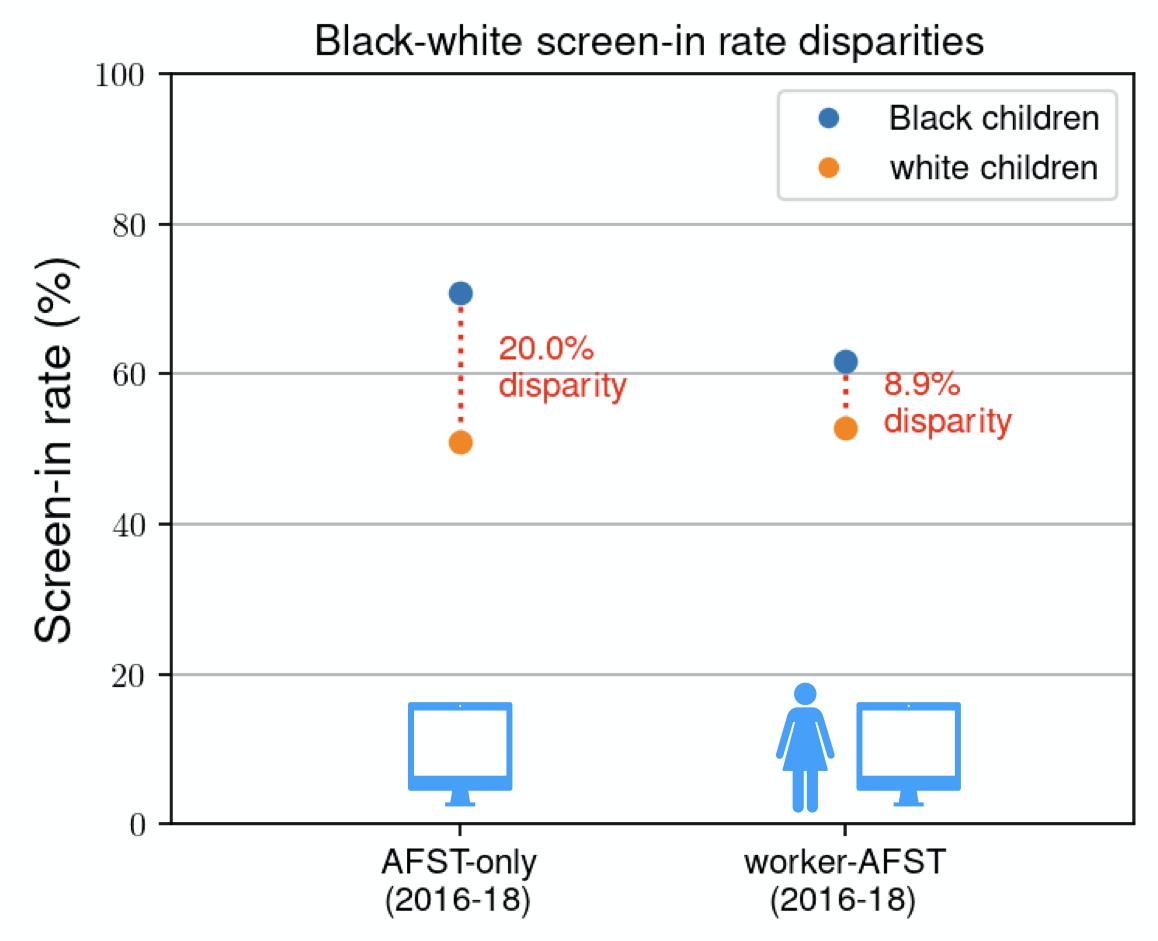}
        \caption{Screen-in rate disparity based on Analysis 1}
        \label{fig:screenin_disparity_analysis1}
    \end{subfigure}%
    ~ 
    \begin{subfigure}[c]{0.5\textwidth}
        \centering
        \includegraphics[width=.96\linewidth]{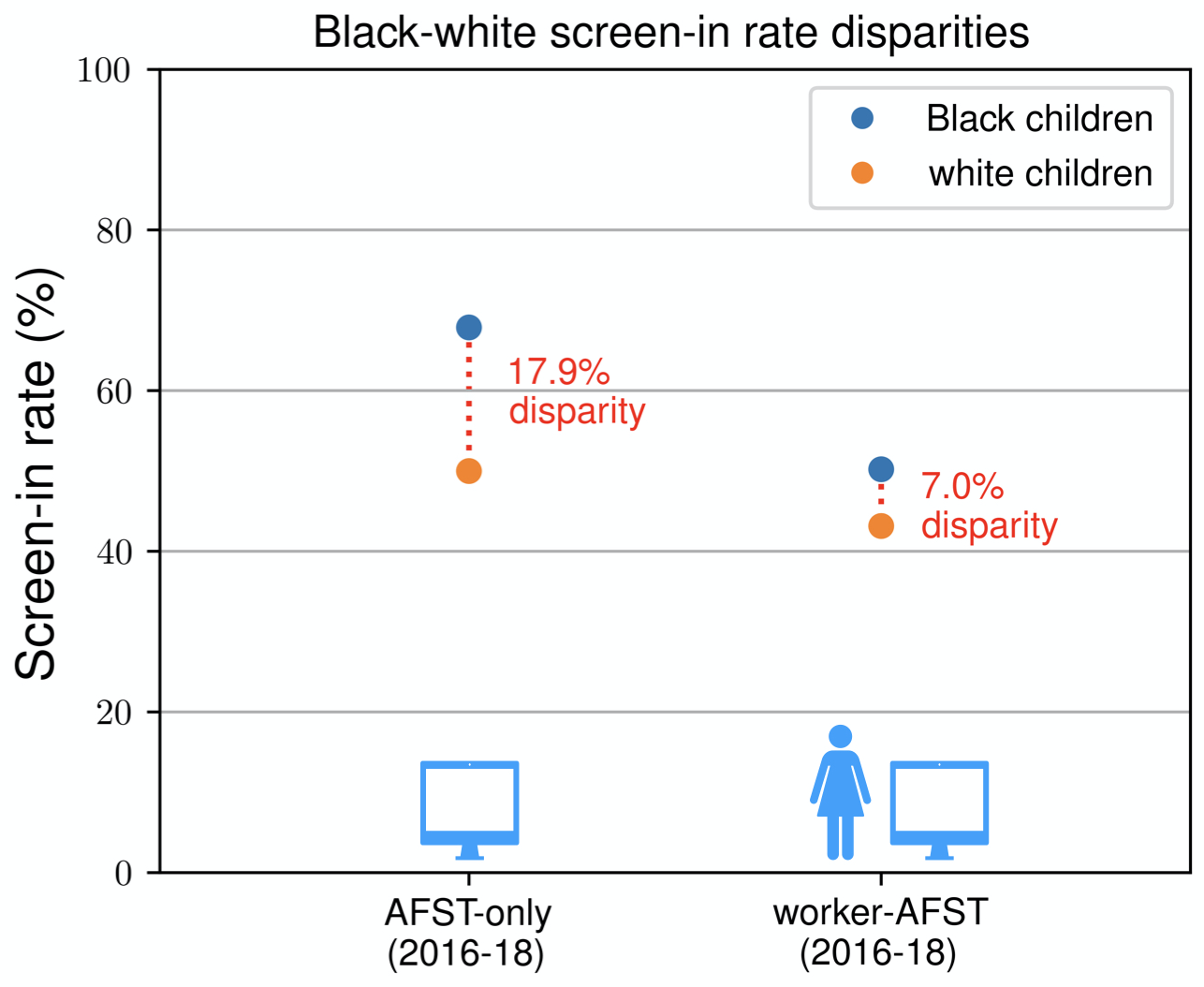}
        \caption{Screen-in rate disparity based on Analysis 2}
        \label{fig:screenin_disparity_analysis2}
    \end{subfigure}
    \caption{Comparison of screen-in rate disparities between Analyses 1 and 2}
    \label{fig:screenin_disparity_comparison}
\end{figure*}

\vspace{-0.5cm}
Although all of the specific numbers we reported in Analysis 1 changed in Analysis 2, our primary conclusions remain the same. Our primary quantitative results are almost identical under both Analyses 1 and 2, even with changes in the data and pre-processing. This lends support to the robustness of our findings that: 1) \textbf{The AFST on its own was more racially disparate than workers, both in terms of screen-in rate and accuracy}; and 2) that \textbf{workers were able to reduce this disparity in the algorithm.} This second result is interesting and surprising, given that child welfare workers are known to make racially disparate decisions without algorithms \cite{Kim2017lifetime}. See Figure~\ref{fig:screenin_disparity_comparison} for a comparison of screen-in rate disparities across Analyses 1 and 2. See Table~\ref{tab:risk_level} for a full comparison of disparities by risk level.

\begin{figure*}[t!]
    \centering
    \begin{subfigure}[c]{0.5\textwidth}
        \centering
        \includegraphics[width=\linewidth]{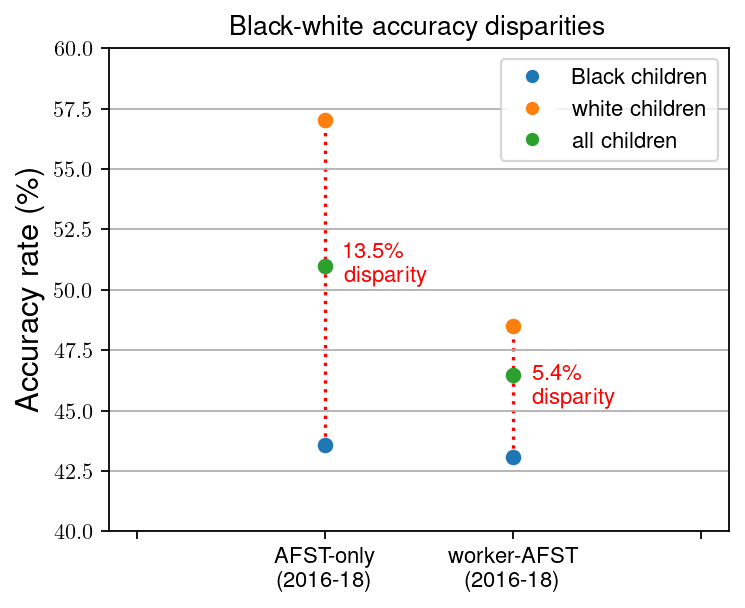}
        \caption{Accuracy rate and disparity based on Analysis 1}
        \label{fig:accuracy_disparity_analysis1}
    \end{subfigure}%
    ~ 
    \begin{subfigure}[c]{0.5\textwidth}
        \centering
        \includegraphics[width=\linewidth]{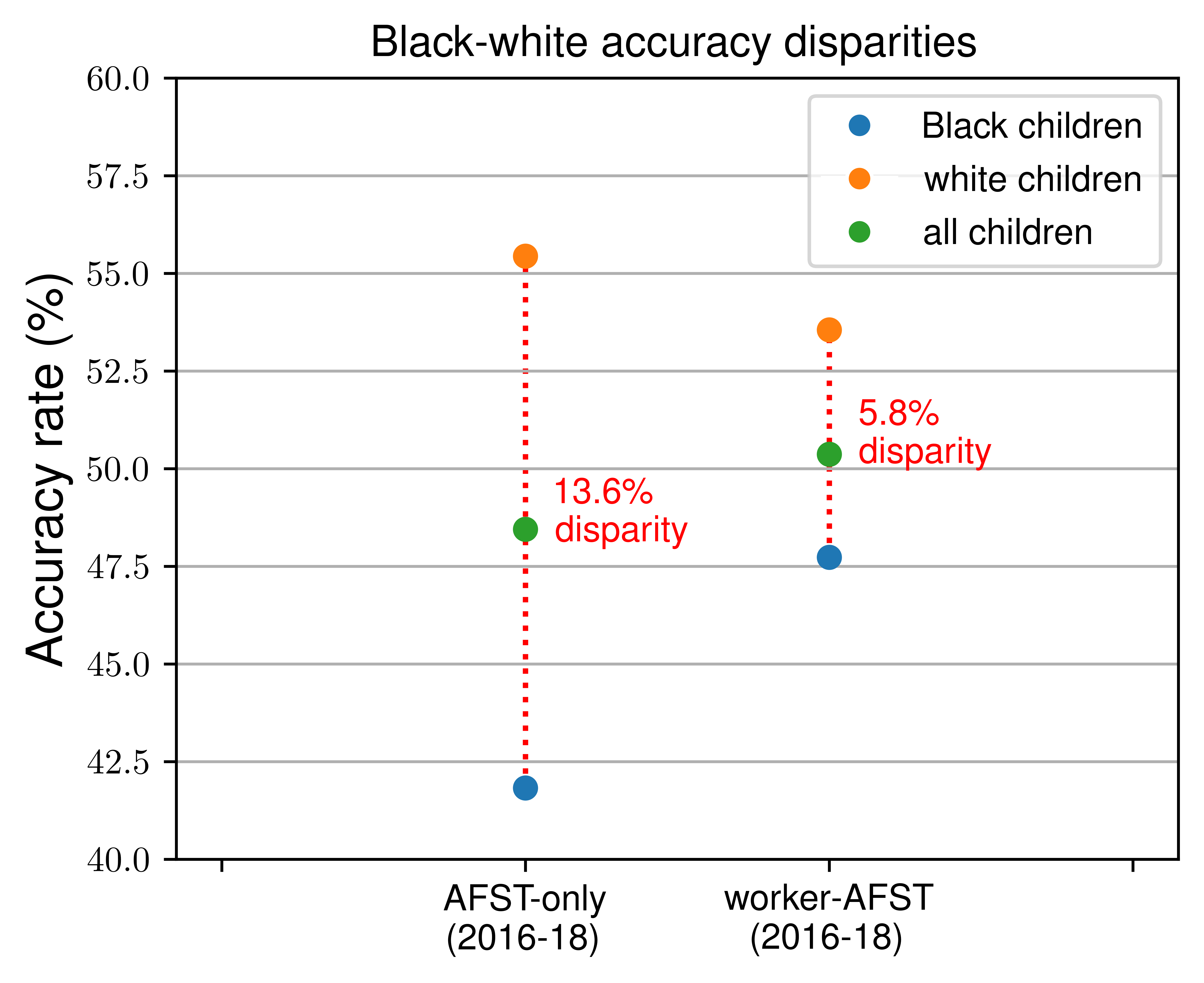}
        \caption{Accuracy rate and disparity based on Analysis 2}
        \label{fig:accuracy_disparity_analysis2}
    \end{subfigure}
    \caption{Comparison of accuracy rates and disparities between Analyses 1 and 2}
    \label{fig:accuracy_disparity_comparison}
\end{figure*}

The two Analyses on disparity in accuracy show similar patterns: AFST-only decisions had higher racial disparity in accuracy than worker-AFST decisions (see Figure~\ref{fig:accuracy_disparity_analysis1} and \ref{fig:accuracy_disparity_analysis2}). However, one aspect of the accuracy comparison that we explicitly mentioned in the paper changed between Analyses: In Analysis 1, we found that overall AFST-only decisions were more accurate than worker-AFST decisions (by 4.5\%). In Analysis 2, we found that overall AFST-only decisions were less accurate than worker-AFST decisions (by 2\%). See Figure~\ref{fig:accuracy_disparity_comparison} on page~\pageref{fig:accuracy_disparity_comparison}. This changes our interpretation of these findings from the original paper. We no longer believe that we can draw clear conclusions based upon the current analyses, about whether the AFST is more accurate than workers or vice versa, even when evaluating performance on the predictive targets used by the AFST itself. It is notable that the results of this comparison changed between Analysis 1 and 2: this highlights the sensitivity of claims about overall accuracy in the AFST to changes in data and pre-processing. However, we \textbf{reiterate the arguments presented in our paper that there are additional reasons to be cautious when interpreting and reporting quantitative results about accuracy and comparisons of accuracy} because of (1) \textbf{the inherent challenges of accuracy measurement in risk assessments for which the predictions affect the predictive targets} (see Section 4.2 of our paper \cite{cheng2022disparities}), and (2) our qualitative finding that workers and the AFST do not agree on prediction outcomes and accuracy measures, meaning that \textbf{evaluating human workers’ performance on the AFST’s benchmarks alone is akin to evaluating a player’s performance at a game they are not actually playing} (see Section 6.2 of our paper \cite{cheng2022disparities}).

Another quantitative result (\textbf{that we did not highlight in the paper due to concerns around confounding factors}) was that the actual screen-in rate increased from before to after the deployment of the AFST. In Analysis 1, we found that, pre-AFST, 52.5\% of Black children and 41.2\% of white children in discretionary referrals were screened in; and post-AFST, 61.8\% of Black children and 52.8\% of white children were screened in. In Analysis 2, we found that, pre-AFST, 50.5\% of Black children and 42.7\% of white children in discretionary referrals were screened in; and post-AFST, 50.2\% of Black children and 43.1\% of white children were screened in. Thus, following Analysis 1, we calculated that the screen-in rate increased about 10\% from pre- to post-AFST, whereas in Analysis 2 there was no noticeable change in the screen-in rate. This indicates that claims about the overall screen-in rate may be more sensitive to changes in data and pre-processing (the design of exclusion criteria).

\xhdr{Summary}
\\
Our main findings were robust to changes in data and pre-processing. Some results changed. However, as these results were less novel or important, we believe they do not significantly alter the primary contributions of our original paper. See the appendix below for a full comparison of Analyses 1 and 2.

\begin{table*}[h]
\scalebox{0.77}{
\begin{tabular}{|l|l|l|l|l|l|l|}
\hline
& \multicolumn{3}{c|}{\cellcolor{tempblue}\textbf{Analysis 1}} & \multicolumn{3}{c|}{\cellcolor{tempviolet}\textbf{Analysis 2}}\\ \hline
\textbf{Risk level}  &  \textbf{All children} &  \textbf{Black children} & \textbf{white children} &  \textbf{All children} &  \textbf{Black children} & \textbf{white children} \\ \hline
All  & 51750   & 26123 (50.5\%) & 21623 (41.8\%) & 35697   & 17292 (48.4\%) & 15596 (43.7\%) \\ \hline
Mandatory & 15182 (29.3\%) & 9639 (36.9\%) & 4863 (22.5\%) & 9431 (26.4\%) & 5618 (32.5\%) & 3242 (20.8\%) \\ \hline
High & 31022 (59.9\%) & 18536 (71.0\%) & 11013 (50.9\%) & 20694 (58.0\%) & 11737 (67.9\%) & 7797 (50.0\%) \\ \hline
Medium & 11778 (22.8\%) & 5208 (19.9\%) & 5653 (26.1\%) & 8391 (23.5\%) & 3778 (21.8\%) & 4025 (25.8\%) \\ \hline
Low & 8950 (17.3\%) & 2379 (9.1\%) & 4957 (22.9\%) & 6612 (18.5\%) & 1777 (10.3\%) & 3774 (24.2\%) \\ \hline
\end{tabular}
}
\caption{Numbers and proportions of (post-AFST) children by risk level and race between Analyses 1 and 2. Percentages are over total children by race, e.g. 1777 Black children labeled Low risk made up 10.3\% of all 17292 Black children referred to CYF from 8/1/16 to 5/13/18.}
\label{tab:risk_level}
\end{table*}

\newpage
\bibliographystyle{ACM-Reference-Format}
\bibliography{reference}


\begin{thebibliography}{5}


\ifx \showCODEN    \undefined \def \showCODEN     #1{\unskip}     \fi
\ifx \showDOI      \undefined \def \showDOI       #1{#1}\fi
\ifx \showISBNx    \undefined \def \showISBNx     #1{\unskip}     \fi
\ifx \showISBNxiii \undefined \def \showISBNxiii  #1{\unskip}     \fi
\ifx \showISSN     \undefined \def \showISSN      #1{\unskip}     \fi
\ifx \showLCCN     \undefined \def \showLCCN      #1{\unskip}     \fi
\ifx \shownote     \undefined \def \shownote      #1{#1}          \fi
\ifx \showarticletitle \undefined \def \showarticletitle #1{#1}   \fi
\ifx \showURL      \undefined \def \showURL       {\relax}        \fi
\providecommand\bibfield[2]{#2}
\providecommand\bibinfo[2]{#2}
\providecommand\natexlab[1]{#1}
\providecommand\showeprint[2][]{arXiv:#2}

\bibitem[\protect\citeauthoryear{Cheng, Stapleton, Kawakami, Sivaraman, Cheng,
  Qing, Perer, Holstein, Wu, and Zhu}{Cheng et~al\mbox{.}}{2022}]%
        {cheng2022disparities}
\bibfield{author}{\bibinfo{person}{Hao-Fei Cheng}, \bibinfo{person}{Logan
  Stapleton}, \bibinfo{person}{Anna Kawakami}, \bibinfo{person}{Venkatesh
  Sivaraman}, \bibinfo{person}{Yanghuidi Cheng}, \bibinfo{person}{Diana Qing},
  \bibinfo{person}{Adam Perer}, \bibinfo{person}{Kenneth Holstein},
  \bibinfo{person}{Zhiwei~Steven Wu}, {and} \bibinfo{person}{Haiyi Zhu}.}
  \bibinfo{year}{2022}\natexlab{}.
\newblock \showarticletitle{How Child Welfare Workers Reduce Racial Disparities
  in Algorithmic Decisions}. In \bibinfo{booktitle}{\emph{Proceedings of the
  2022 CHI Conference on Human Factors in Computing Systems}}.
\newblock
\urldef\tempurl%
\url{https://dl.acm.org/doi/abs/10.1145/3491102.3501831}
\showURL{%
\tempurl}


\bibitem[\protect\citeauthoryear{De-Arteaga, Fogliato, and
  Chouldechova}{De-Arteaga et~al\mbox{.}}{2020}]%
        {De-Arteaga2020}
\bibfield{author}{\bibinfo{person}{Maria De-Arteaga}, \bibinfo{person}{Riccardo
  Fogliato}, {and} \bibinfo{person}{Alexandra Chouldechova}.}
  \bibinfo{year}{2020}\natexlab{}.
\newblock \showarticletitle{A case for humans-in-the-loop: Decisions in the
  presence of erroneous algorithmic scores}. In
  \bibinfo{booktitle}{\emph{Proceedings of the 2020 CHI Conference on Human
  Factors in Computing Systems}}. \bibinfo{pages}{1--12}.
\newblock


\bibitem[\protect\citeauthoryear{Goldhaber-Fiebert and
  Prince}{Goldhaber-Fiebert and Prince}{2019}]%
        {goldhaber2019impact}
\bibfield{author}{\bibinfo{person}{Jeremy~D Goldhaber-Fiebert} {and}
  \bibinfo{person}{Lea Prince}.} \bibinfo{year}{2019}\natexlab{}.
\newblock \showarticletitle{Impact evaluation of a predictive risk modeling
  tool for Allegheny county’s child welfare office}.
\newblock \bibinfo{journal}{\emph{Pittsburgh: Allegheny County}}
  (\bibinfo{year}{2019}).
\newblock


\bibitem[\protect\citeauthoryear{Kim, Wildeman, Jonson-Reid, and Drake}{Kim
  et~al\mbox{.}}{2017}]%
        {Kim2017lifetime}
\bibfield{author}{\bibinfo{person}{Hyunil Kim}, \bibinfo{person}{Christopher
  Wildeman}, \bibinfo{person}{Melissa Jonson-Reid}, {and}
  \bibinfo{person}{Brett Drake}.} \bibinfo{year}{2017}\natexlab{}.
\newblock \showarticletitle{Lifetime Prevalence of Investigating Child
  Maltreatment Among US Children}.
\newblock \bibinfo{journal}{\emph{American Journal of Public Health}}
  \bibinfo{volume}{107}, \bibinfo{number}{2} (\bibinfo{year}{2017}),
  \bibinfo{pages}{274--280}.
\newblock
\urldef\tempurl%
\url{https://doi.org/10.2105/AJPH.2016.303545}
\showDOI{\tempurl}


\bibitem[\protect\citeauthoryear{Vaithianathan, Jiang, Maloney, Nand, and
  Putnam-Hornstein}{Vaithianathan et~al\mbox{.}}{2017}]%
        {vaithianathan2017}
\bibfield{author}{\bibinfo{person}{Rhema Vaithianathan}, \bibinfo{person}{Nan
  Jiang}, \bibinfo{person}{Tim Maloney}, \bibinfo{person}{Parma Nand}, {and}
  \bibinfo{person}{Emily Putnam-Hornstein}.} \bibinfo{year}{2017}\natexlab{}.
\newblock \bibinfo{title}{Developing Predictive Risk Models to Support Child
  Maltreatment Hotline Screening Decisions}.
\newblock
\newblock
\urldef\tempurl%
\url{https://www.alleghenycountyanalytics.us/wp-content/uploads/2017/04/Developing-Predictive-Risk-Models-package-with-cover-1-to-post-1.pdf}
\showURL{%
\tempurl}


\end{thebibliography}

\appendix


\section{Detailed Comparisons of Analyses 1 and 2}

\begin{table*}[h]
    \centering
    \includegraphics[width=.9\linewidth]{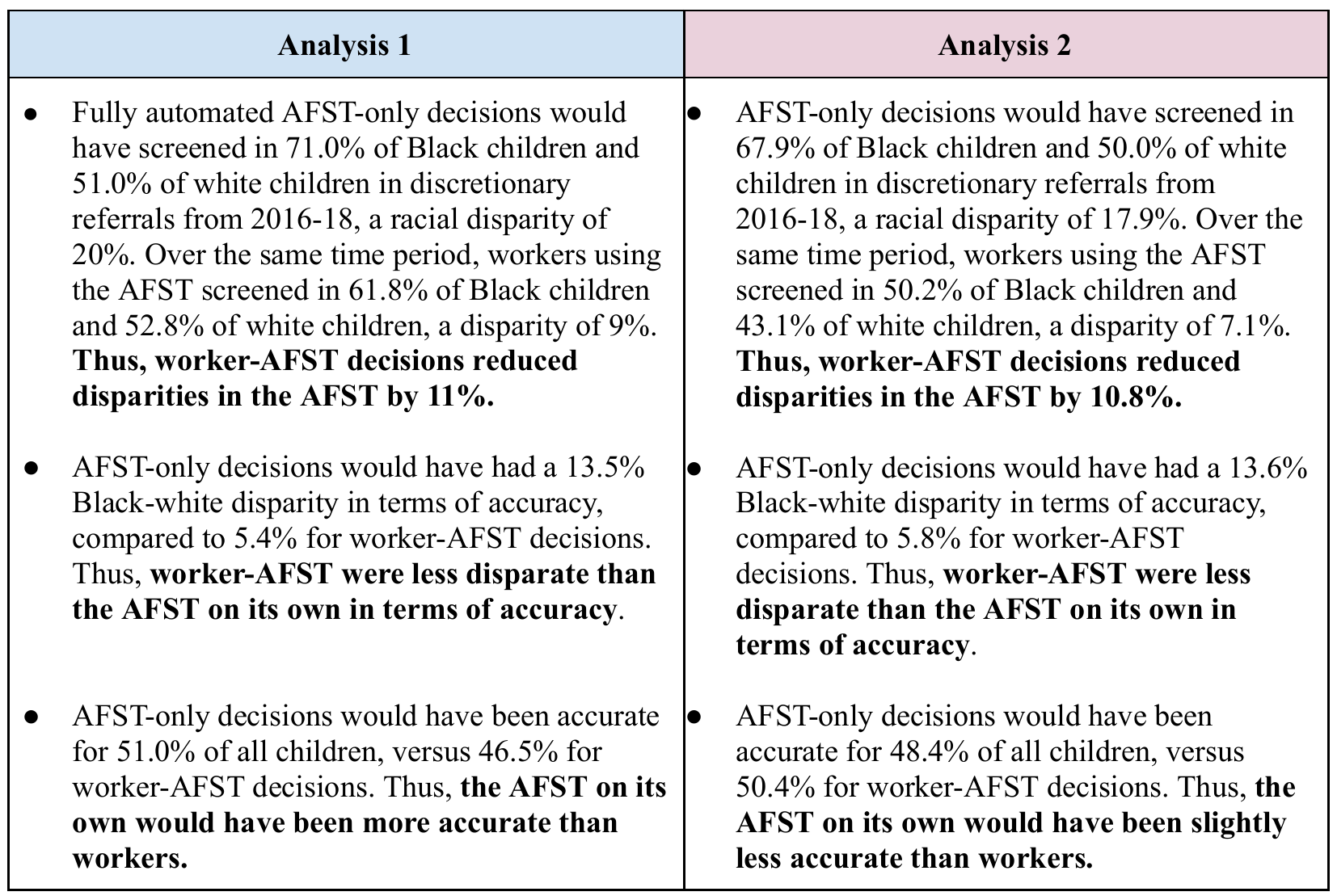}
    \caption{Comparison of quantitative findings between Analyses 1 and 2}
    \label{tab:findings_comparison}
\end{table*}

\begin{table*}[h]
    \centering
    \includegraphics[width=\linewidth]{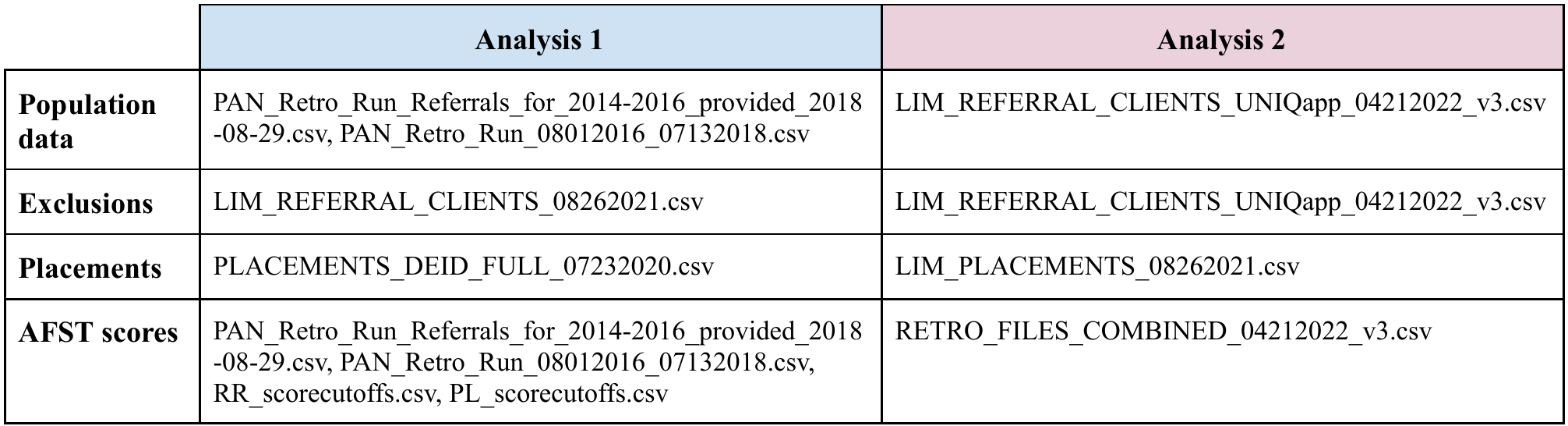}
    \caption{Comparison of data used between Analyses 1 and 2}
    \label{tab:data_comparison}
\end{table*}

\begin{table*}[h]
    \centering
    \includegraphics[width=\linewidth]{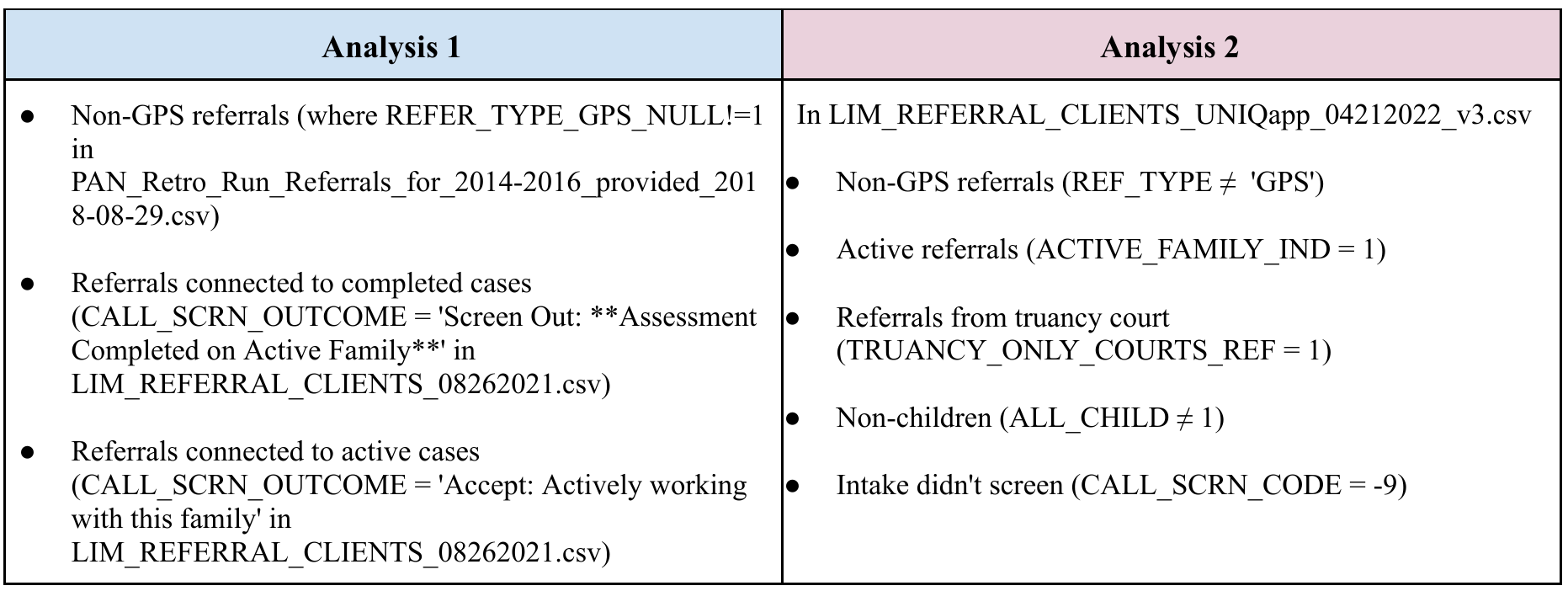}
    \caption{Comparison of data exclusions steps between Analyses 1 and 2}
    \label{tab:exclusions_comparison}
\end{table*}

\begin{table*}[h]
    \centering
    \includegraphics[width=\linewidth]{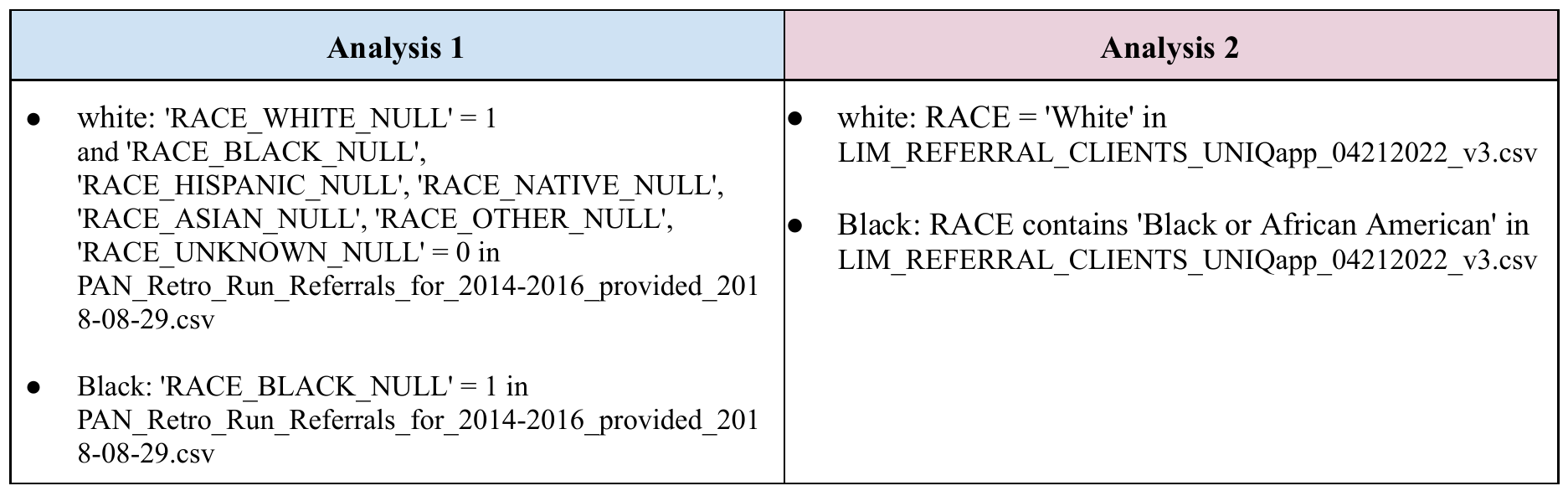}
    \caption{Comparison of race (Black and white) coding between Analyses 1 and 2}
    \label{tab:racecoding_comparison}
\end{table*}


\pagebreak

\section{Comparisons of all figures in the paper \cite{cheng2022disparities} based on Analyses 1 and 2}

\begin{figure*}
    \centering
    \caption{Comparison of worker-AFST compliance rates (Figure 4 in the paper \cite{cheng2022disparities}) based on Analyses 1 and 2}
    \label{fig:compliance}
    \begin{subfigure}[c]{0.5\textwidth}
        \centering
         \includegraphics[width=\linewidth]{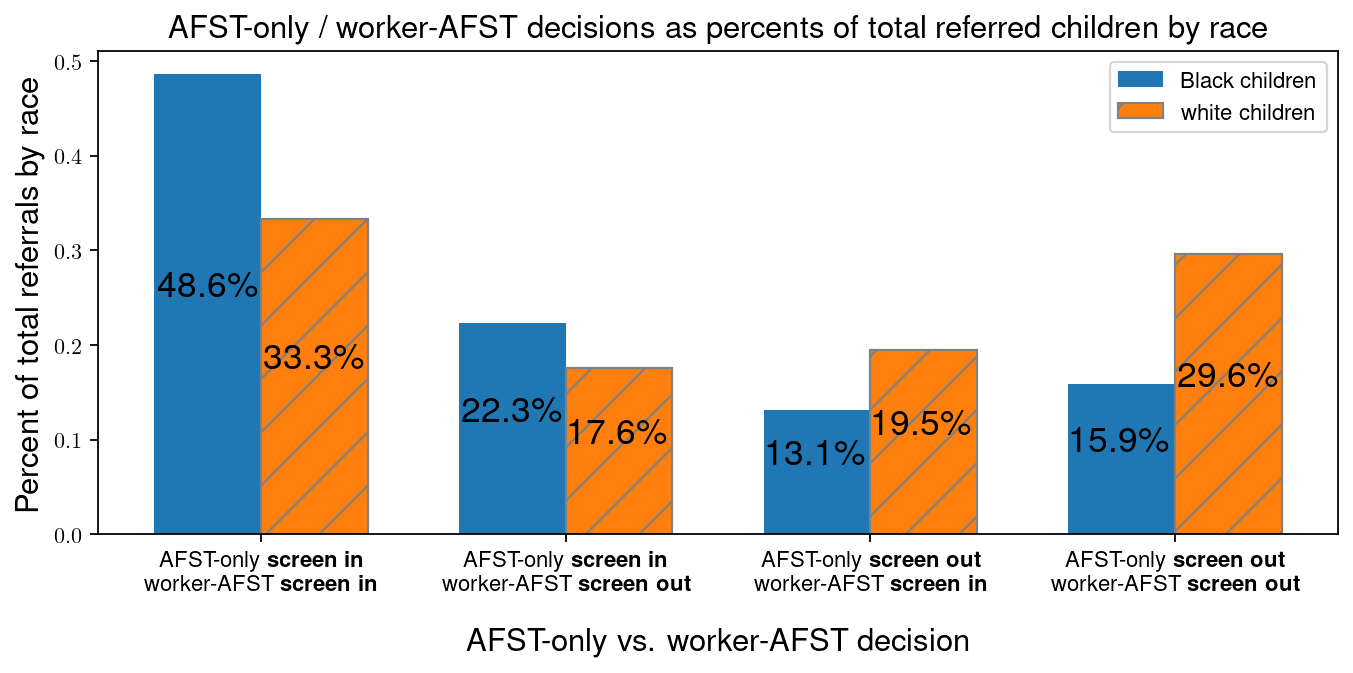}
        \caption{Figure 4 in the paper \cite{cheng2022disparities} based on Analysis 1}
        \label{fig:compliance_analysis1}
    \end{subfigure}%
    ~ 
    \begin{subfigure}[c]{0.5\textwidth}
        \centering
        \includegraphics[width=\linewidth]{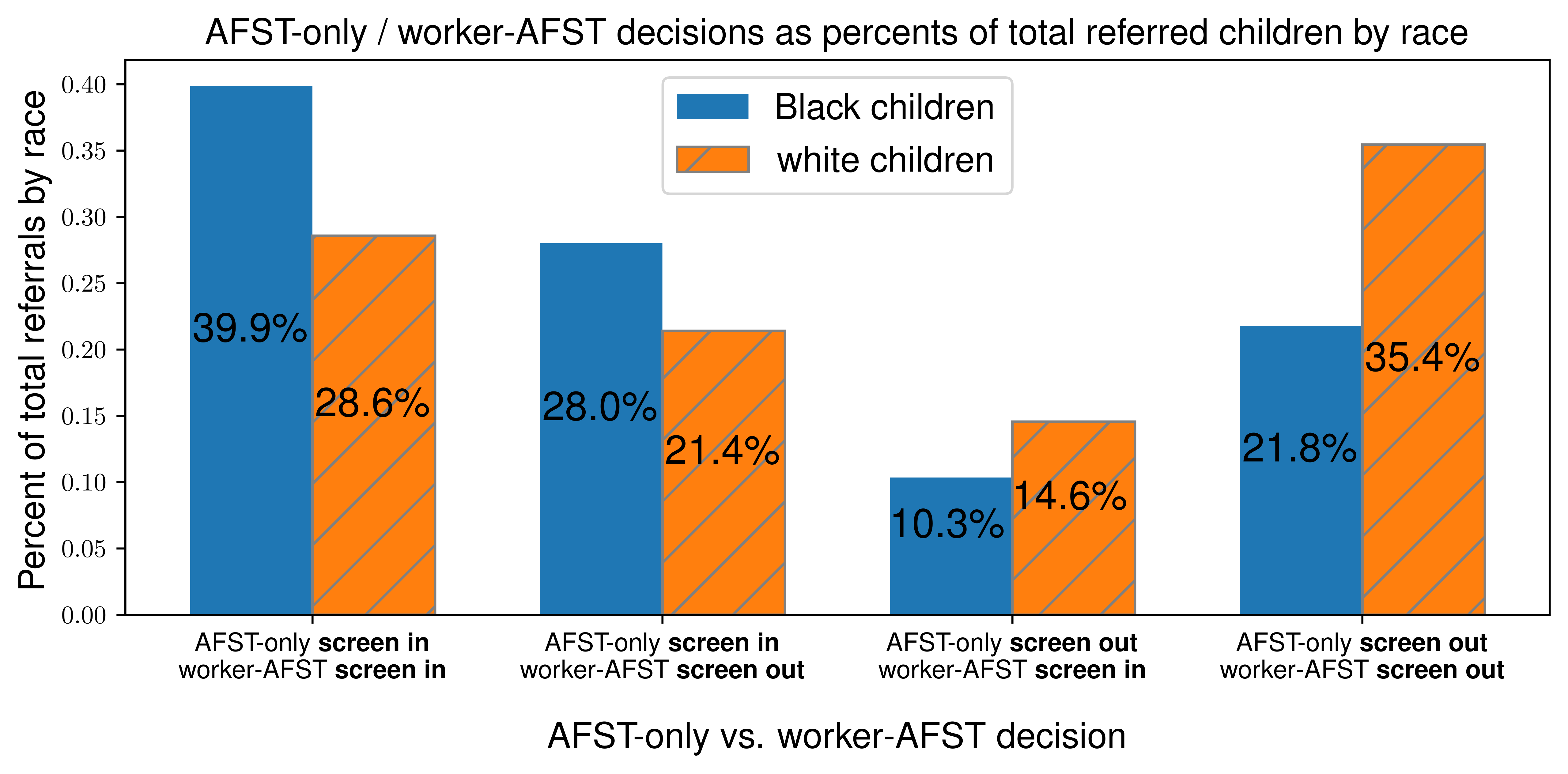}
        \caption{Figure 4 in the paper \cite{cheng2022disparities} based on Analysis 2}
        \label{fig:compliance_analysis2}
    \end{subfigure}
\end{figure*}

\begin{figure*}
    \centering
    \caption{Comparisons of common group fairness metrics (Figure 6 in the paper \cite{cheng2022disparities}) based on Analysis 1}
    \label{fig:human-ai_comparison_analysis1}
    \begin{subfigure}[t]{0.48\textwidth}
        \centering
        \includegraphics[width=\linewidth]{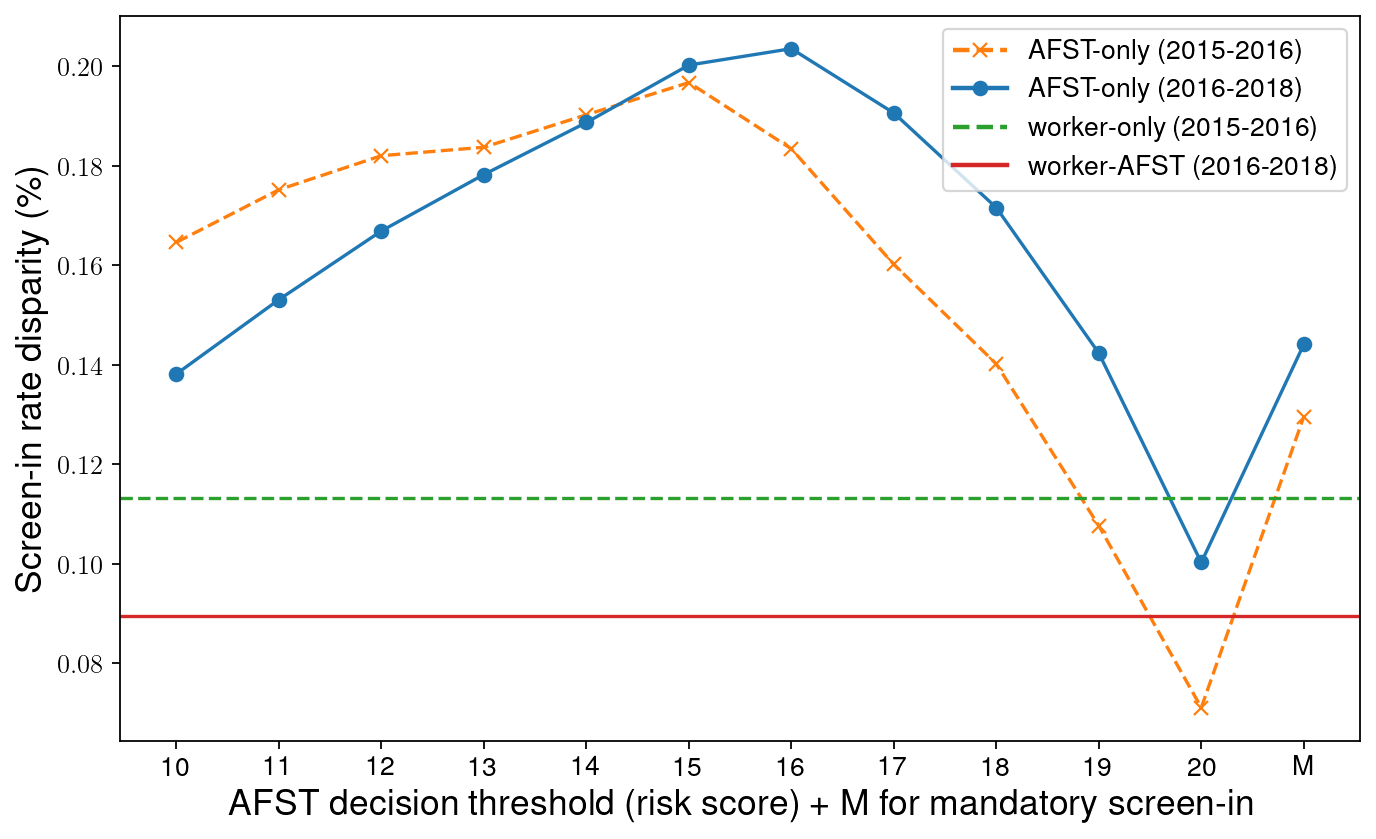} 
        \caption{Screen-in rate disparity} \label{fig:human-ai_comparison_statistical_parity_analysis1}
    \end{subfigure}
    \hfill
    \begin{subfigure}[t]{0.48\textwidth}
        \centering
        \includegraphics[width=\linewidth]{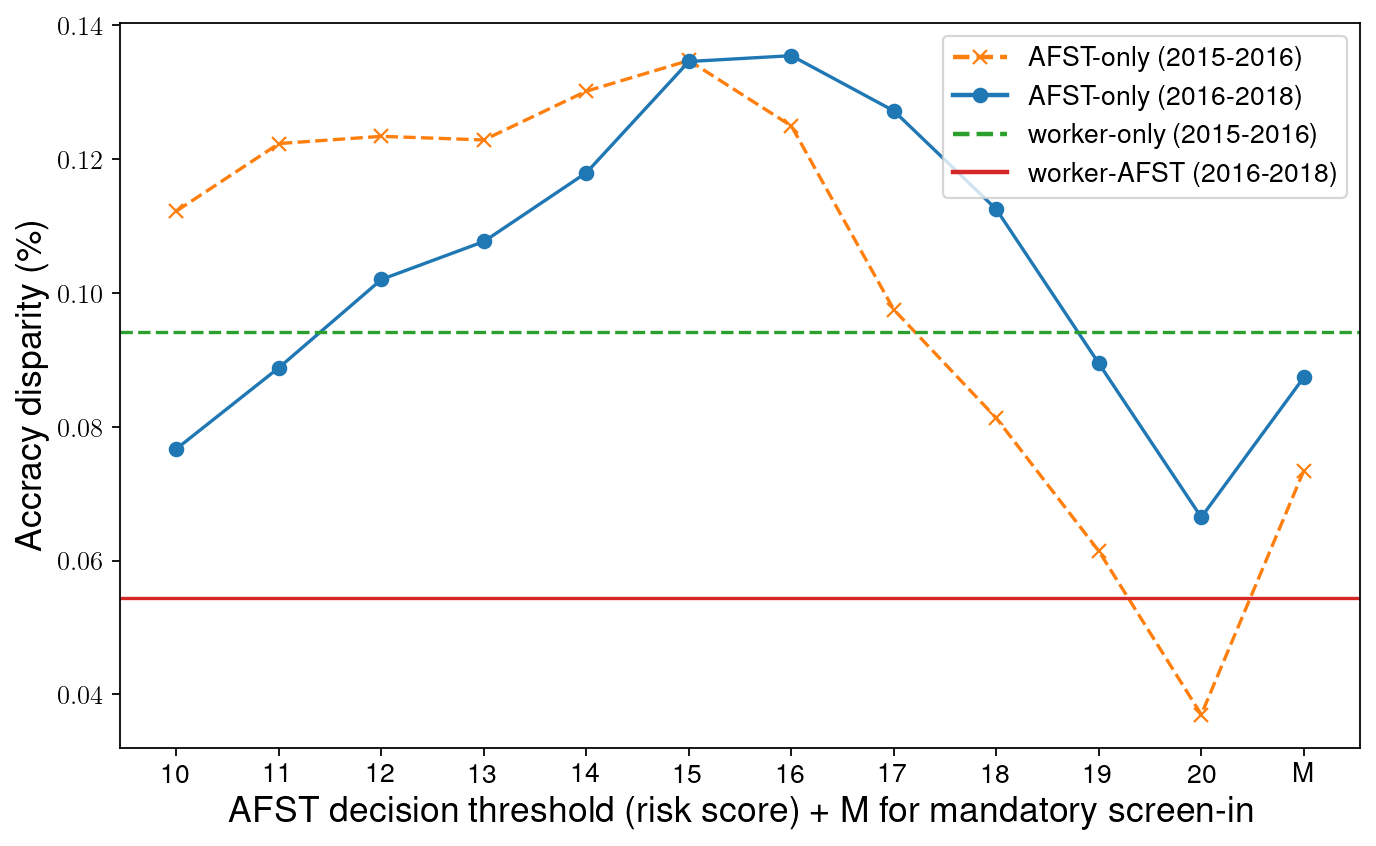} 
        \caption{Accuracy disparity} \label{fig:human-ai_comparison_accuracy_parity_analysis1}
    \end{subfigure}
    \vspace{1cm}
    \begin{subfigure}[t]{0.48\textwidth}
        \centering
        \includegraphics[width=\linewidth]{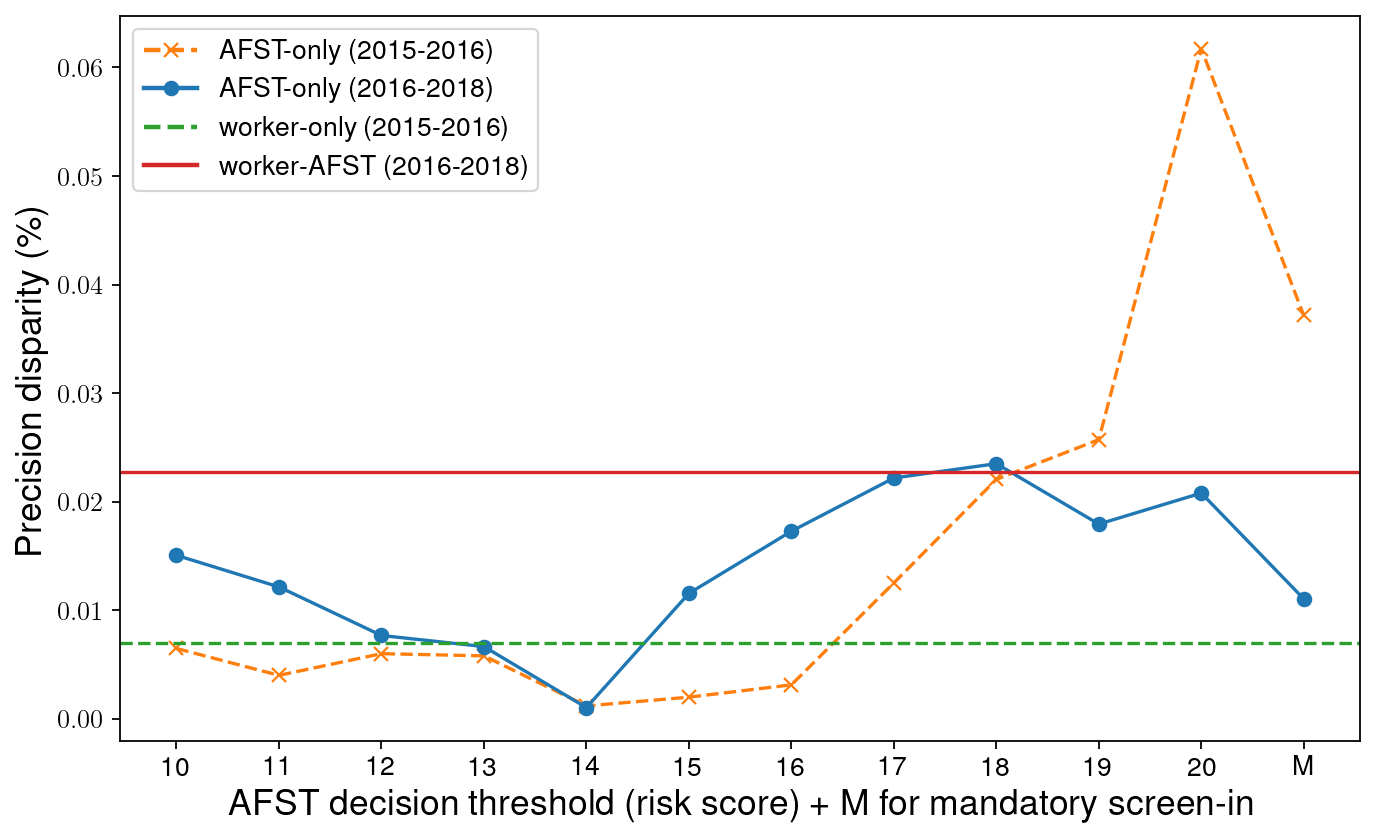} 
        \caption{Precision rate disparity} \label{fig:human-ai_comparison_predictive_parity_analysis1}
    \end{subfigure}
    \hfill
    \begin{subfigure}[t]{0.48\textwidth}
        \centering
        \includegraphics[width=\linewidth]{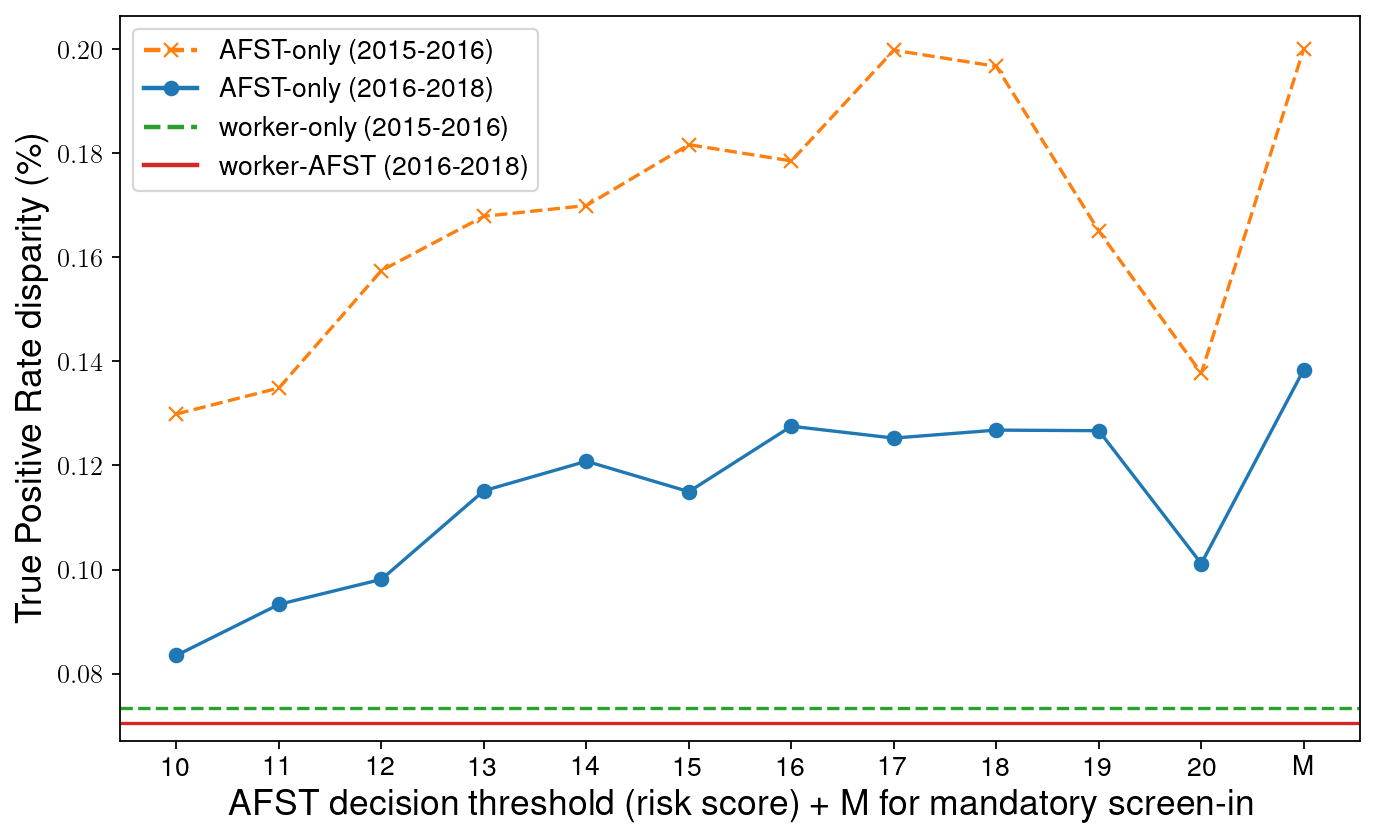} 
        \caption{True positive rate disparity} \label{fig:human-ai_comparison_TPR_parity_analysis1}
    \end{subfigure}
    \begin{subfigure}[t]{0.48\textwidth}
        \centering
        \includegraphics[width=\linewidth]{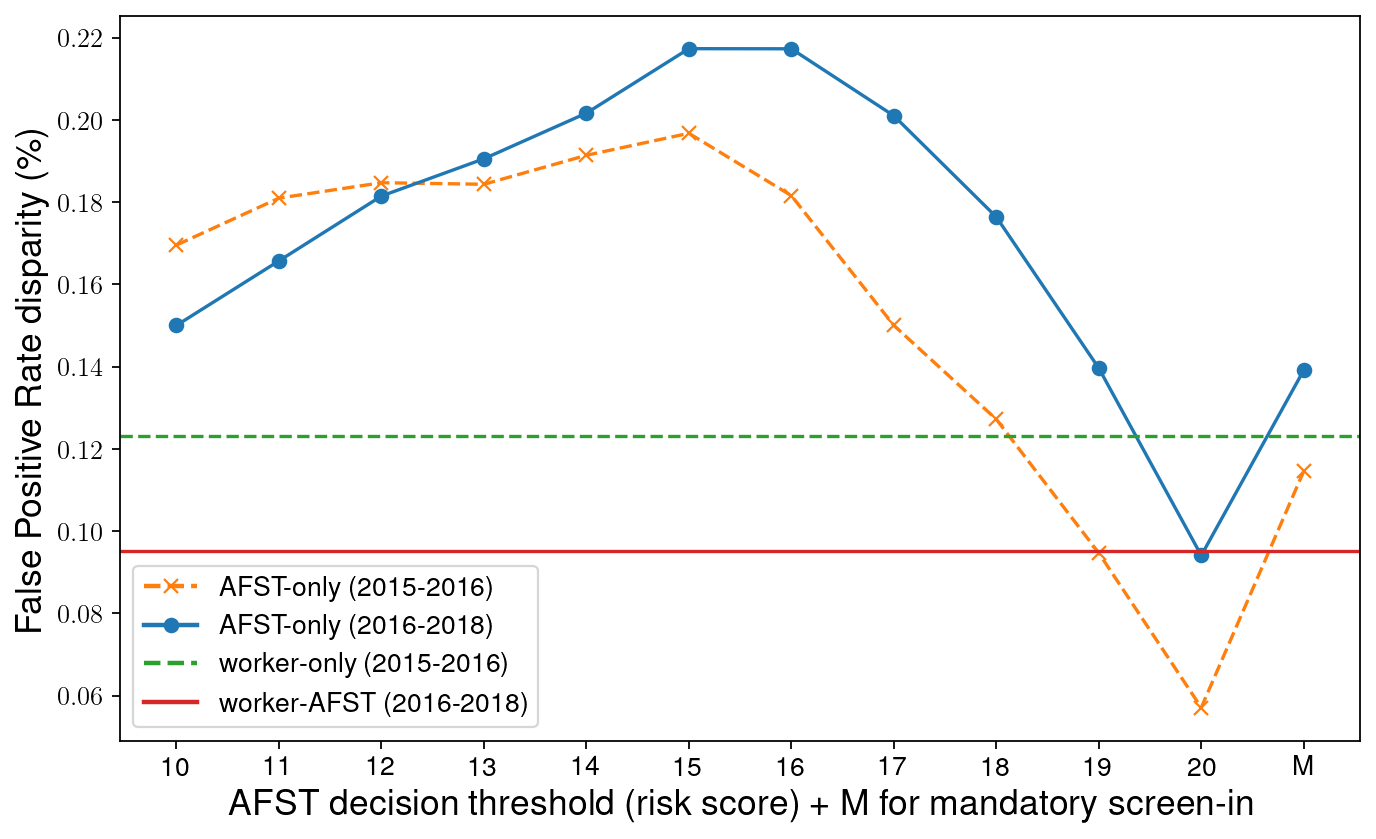} 
        \caption{False positive rate disparity} \label{fig:human-ai_comparison_FPR_parity_analysis1}
    \end{subfigure}
\end{figure*}

\begin{figure*}
    \centering
    \caption{Comparisons of common group fairness metrics (Figure 6 in the paper \cite{cheng2022disparities}) based on Analysis 2}
    \label{fig:human-ai_comparison_analysis2}
    \begin{subfigure}[t]{0.48\textwidth}
        \centering
        \includegraphics[width=\linewidth]{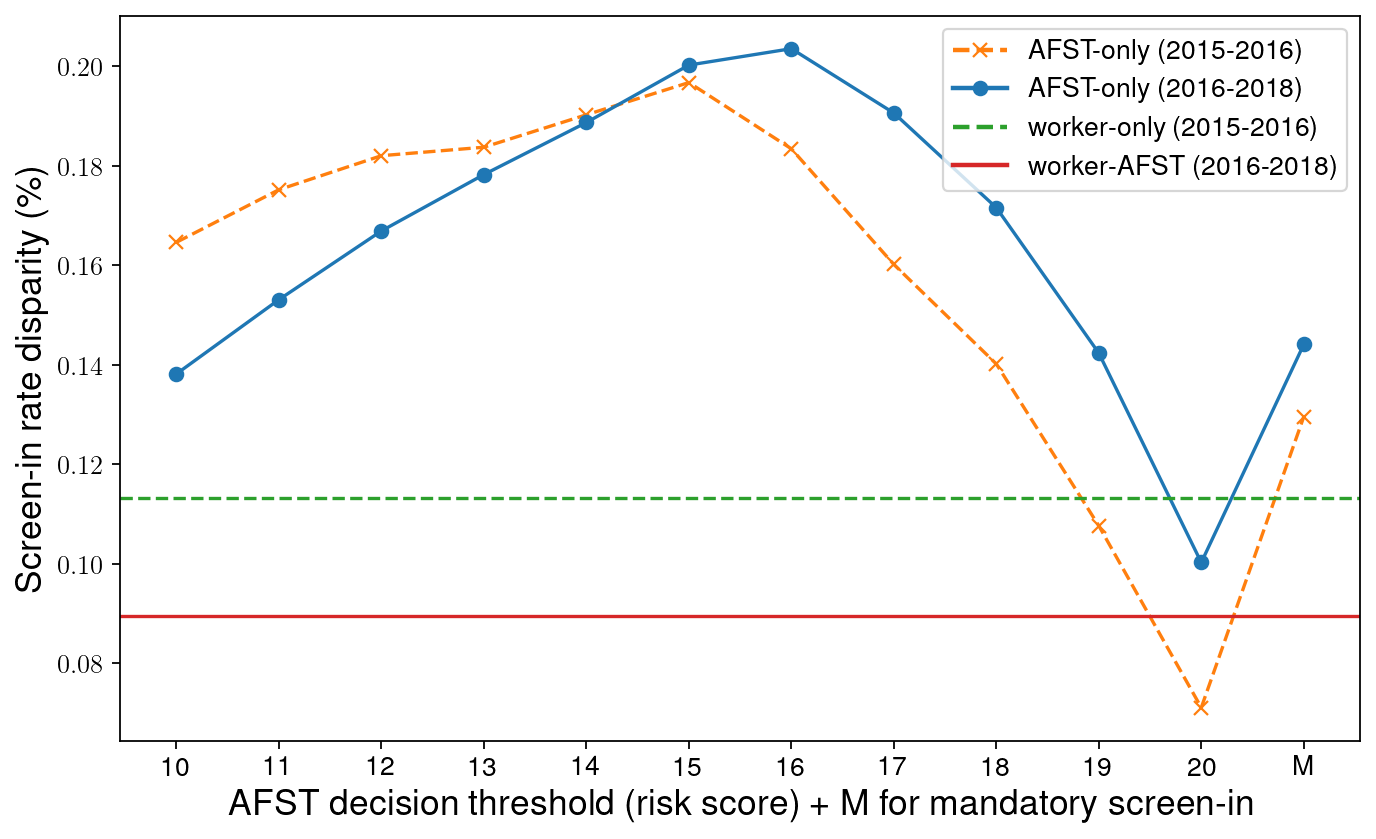} 
        \caption{Screen-in rate disparity} \label{fig:human-ai_comparison_statistical_parity_analysis2}
    \end{subfigure}
    \hfill
    \begin{subfigure}[t]{0.48\textwidth}
        \centering
        \includegraphics[width=\linewidth]{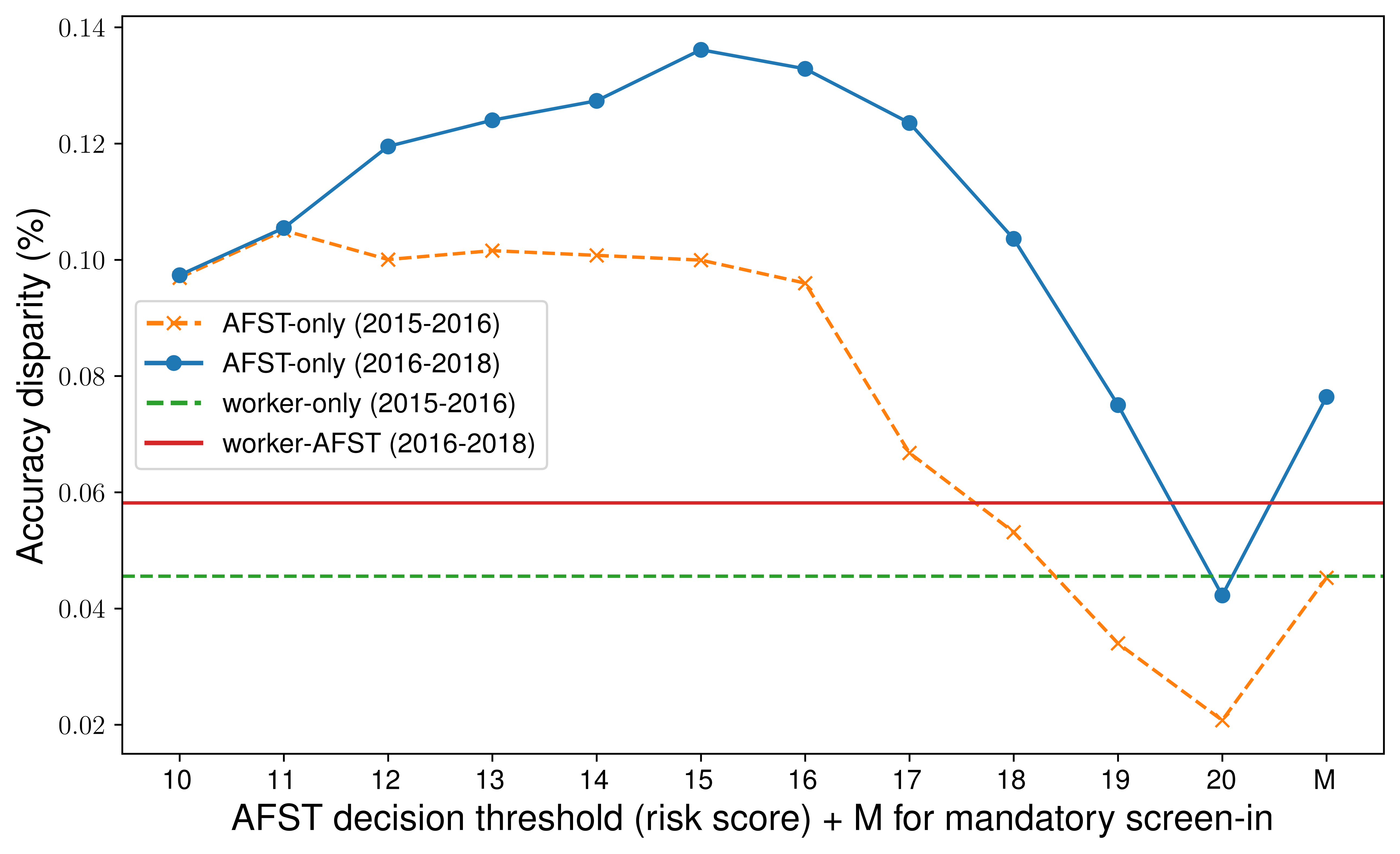} 
        \caption{Accuracy disparity} \label{fig:human-ai_comparison_accuracy_parity_analysis2}
    \end{subfigure}
    \vspace{1cm}
    \begin{subfigure}[t]{0.48\textwidth}
        \centering
        \includegraphics[width=\linewidth]{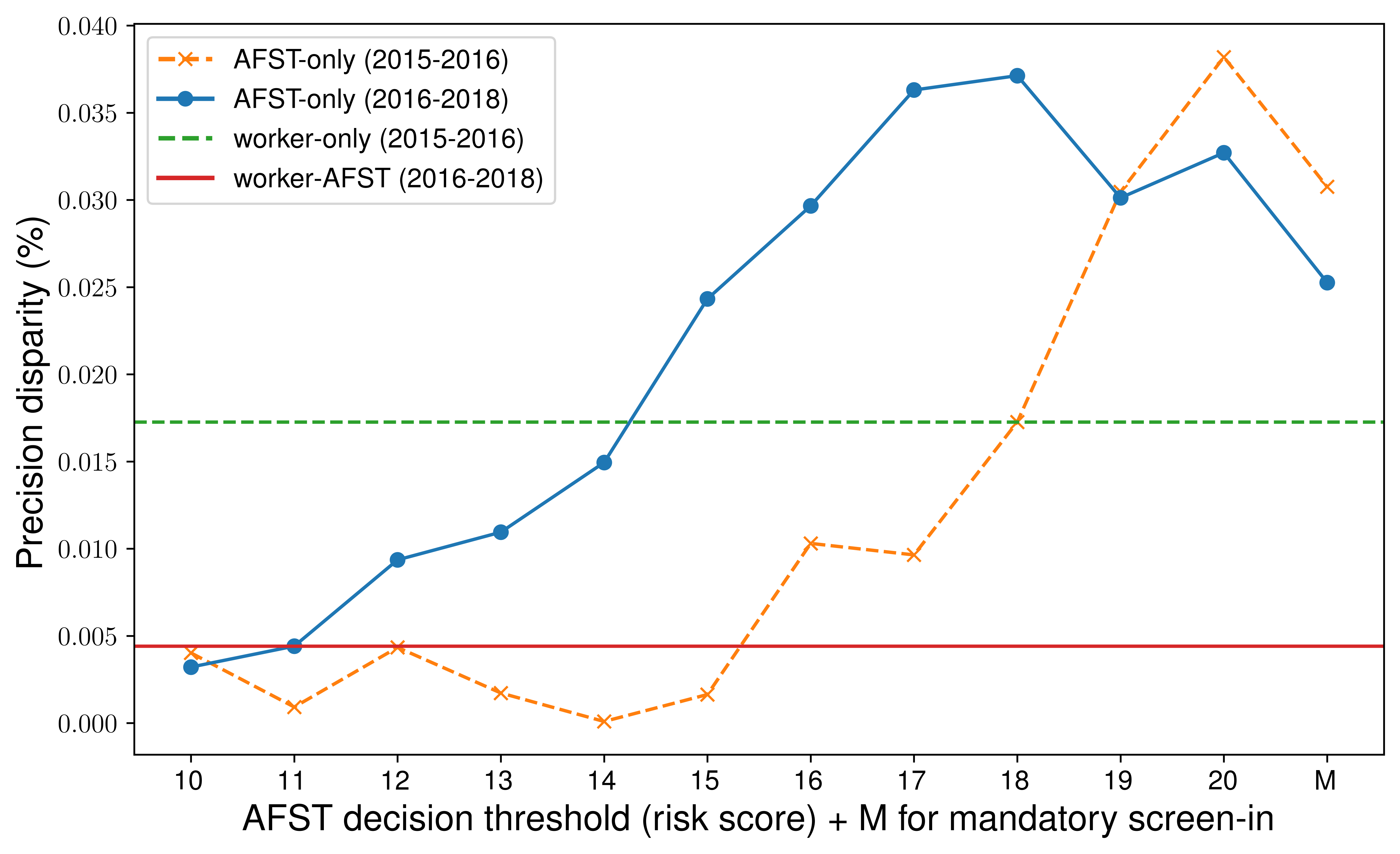} 
        \caption{Precision rate disparity} \label{fig:human-ai_comparison_predictive_parity_analysis2}
    \end{subfigure}
    \hfill
    \begin{subfigure}[t]{0.48\textwidth}
        \centering
        \includegraphics[width=\linewidth]{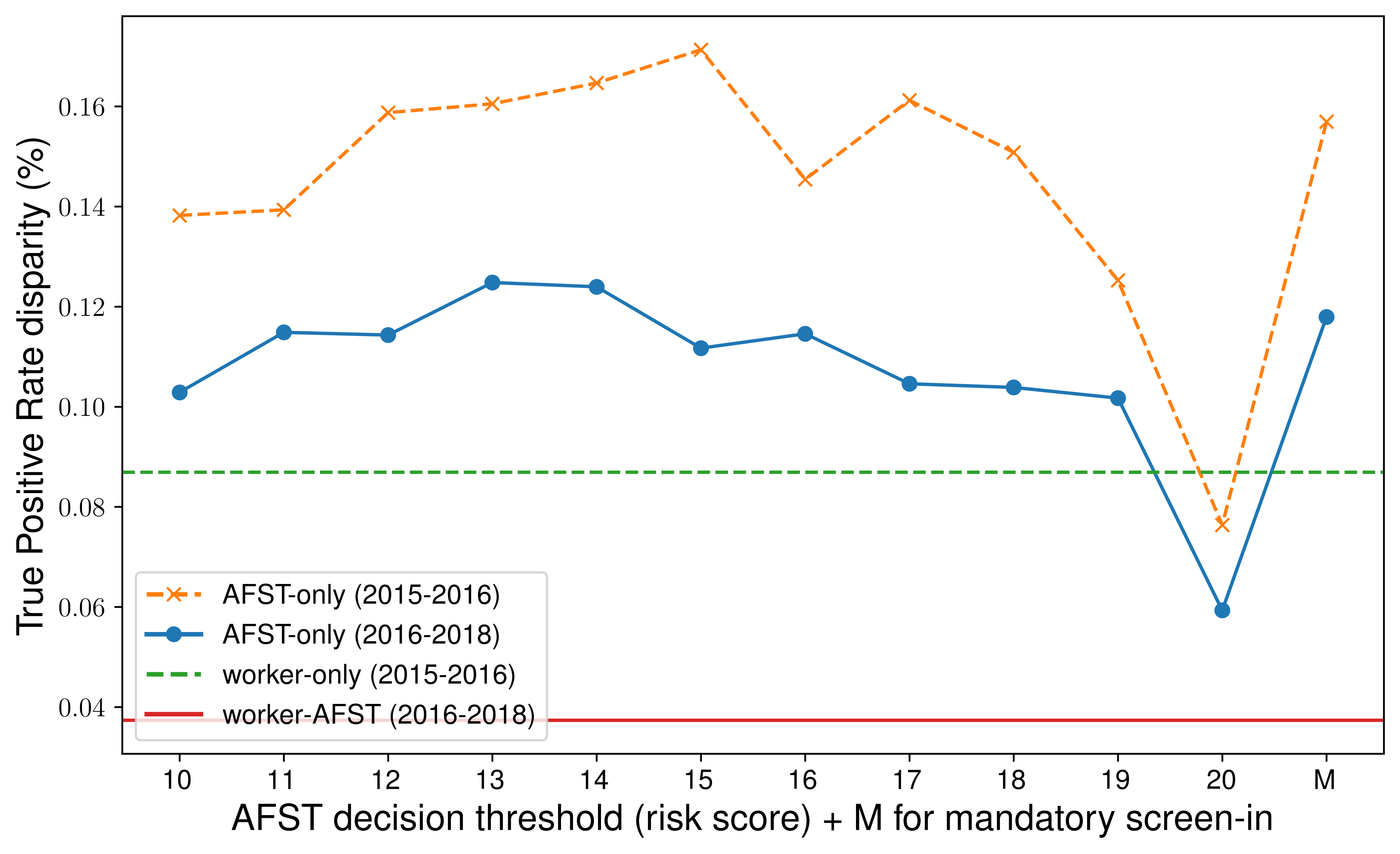} 
        \caption{True positive rate disparity} \label{fig:human-ai_comparison_TPR_parity_analysis2}
    \end{subfigure}
    \begin{subfigure}[t]{0.48\textwidth}
        \centering
        \includegraphics[width=\linewidth]{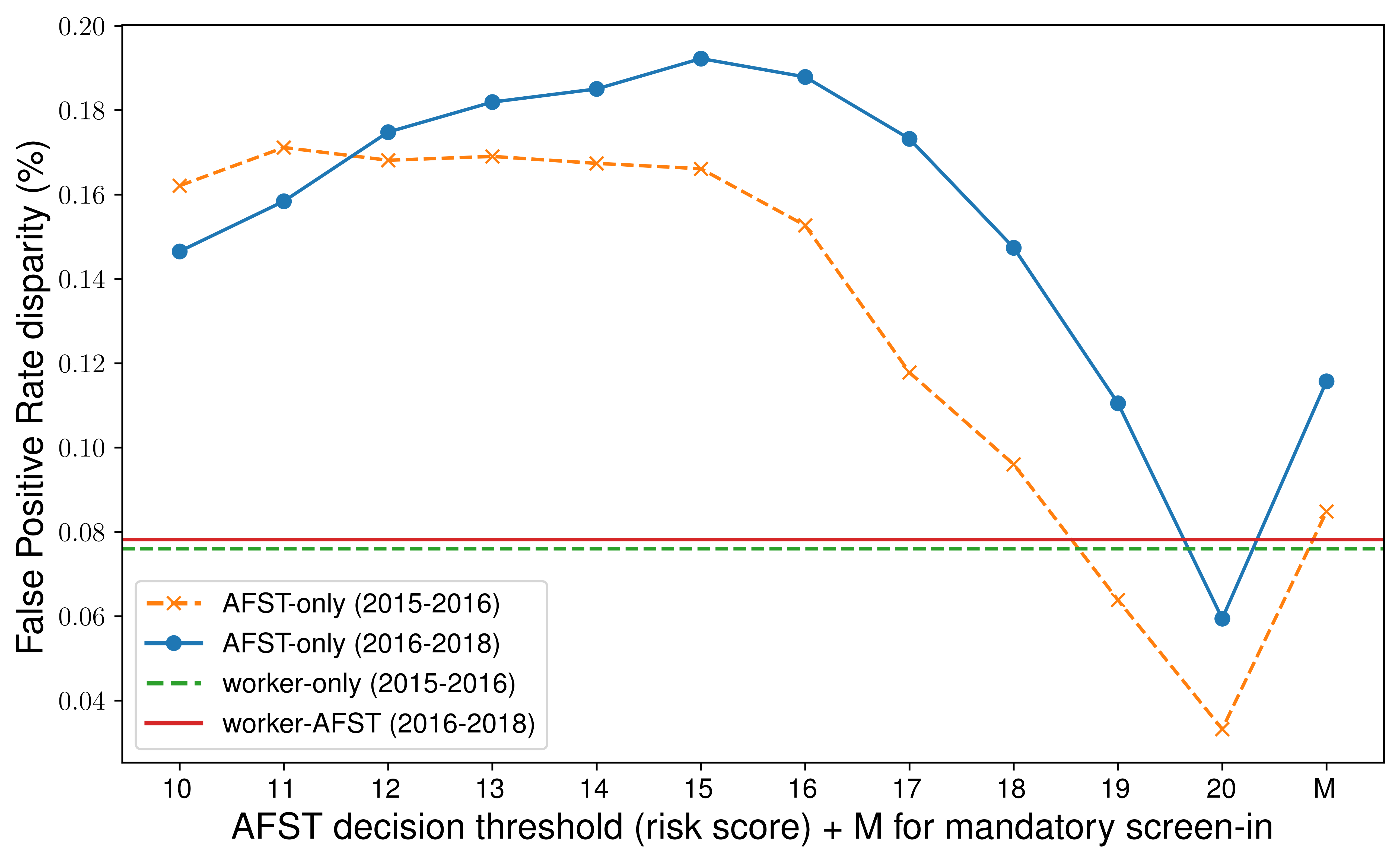} 
        \caption{False positive rate disparity} \label{fig:human-ai_comparison_FPR_parity_analysis2}
    \end{subfigure}
\end{figure*}

\begin{figure*}[h]
    \centering
    \caption{Comparisons of decision outcomes between worker-AFST and AFST-only decisions (Figure 7 in the paper \cite{cheng2022disparities}) based on Analysis 1}
    \label{fig:per_threshold_analysis1}
    \begin{subfigure}[t]{0.48\textwidth}
        \centering
        \includegraphics[width=\linewidth]{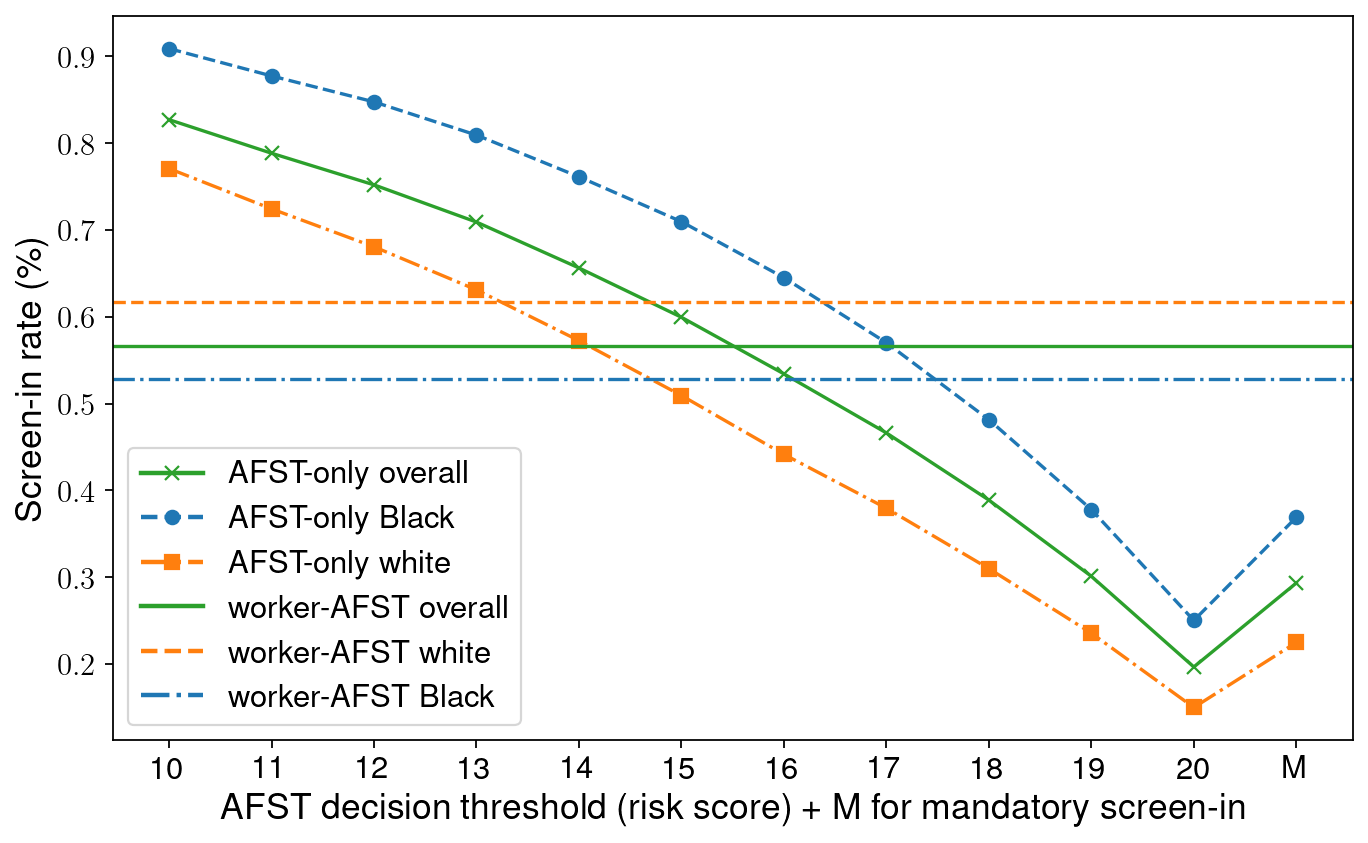} 
        \caption{Screen-in rate} \label{fig:per_threshold_screen_in_rate_analysis1}
    \end{subfigure}
    \hfill
    \begin{subfigure}[t]{0.48\textwidth}
        \centering
        \includegraphics[width=\linewidth]{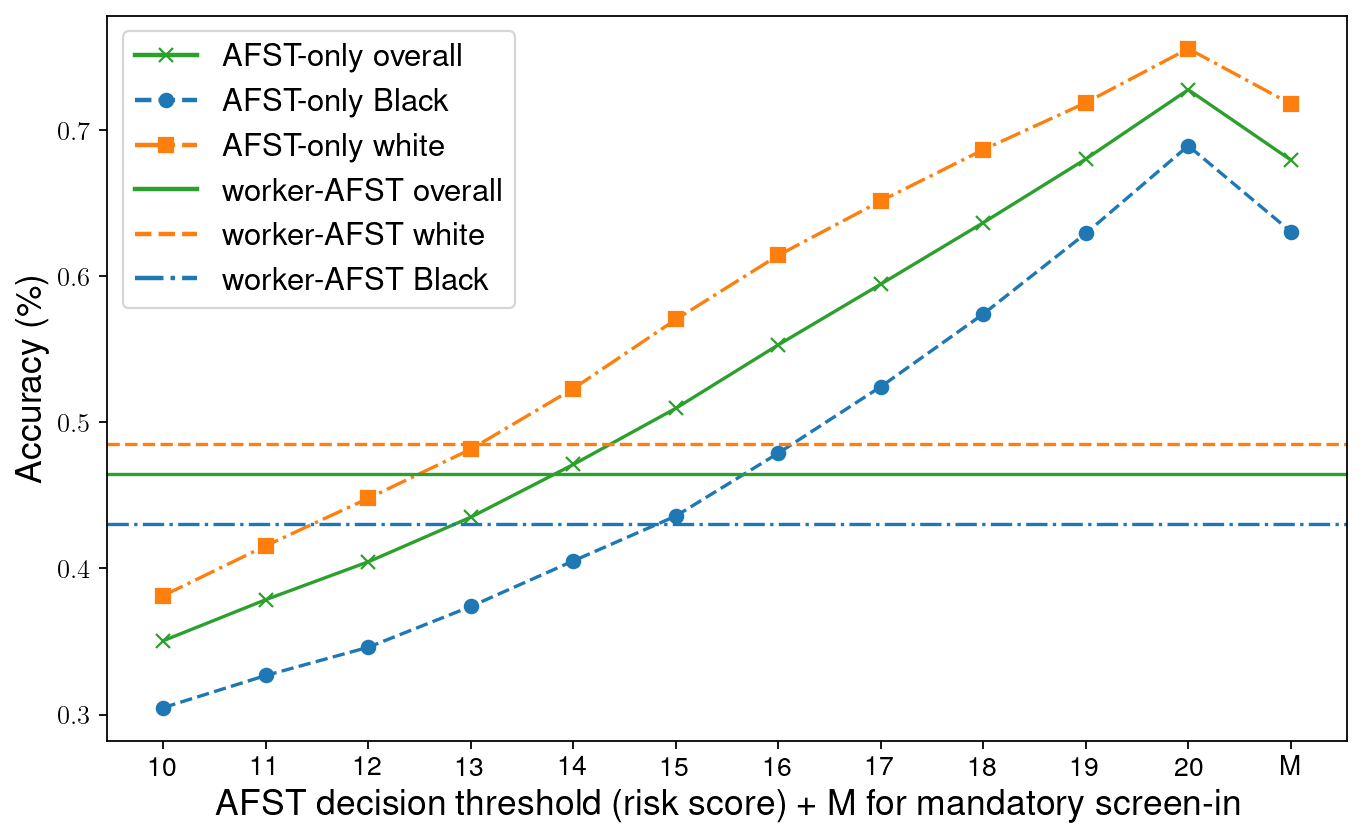} 
        \caption{Accuracy} \label{fig:per_threshold_accuracy_analysis1}
    \end{subfigure}
    \vspace{1cm}
    \begin{subfigure}[t]{0.48\textwidth}
        \centering
        \includegraphics[width=\linewidth]{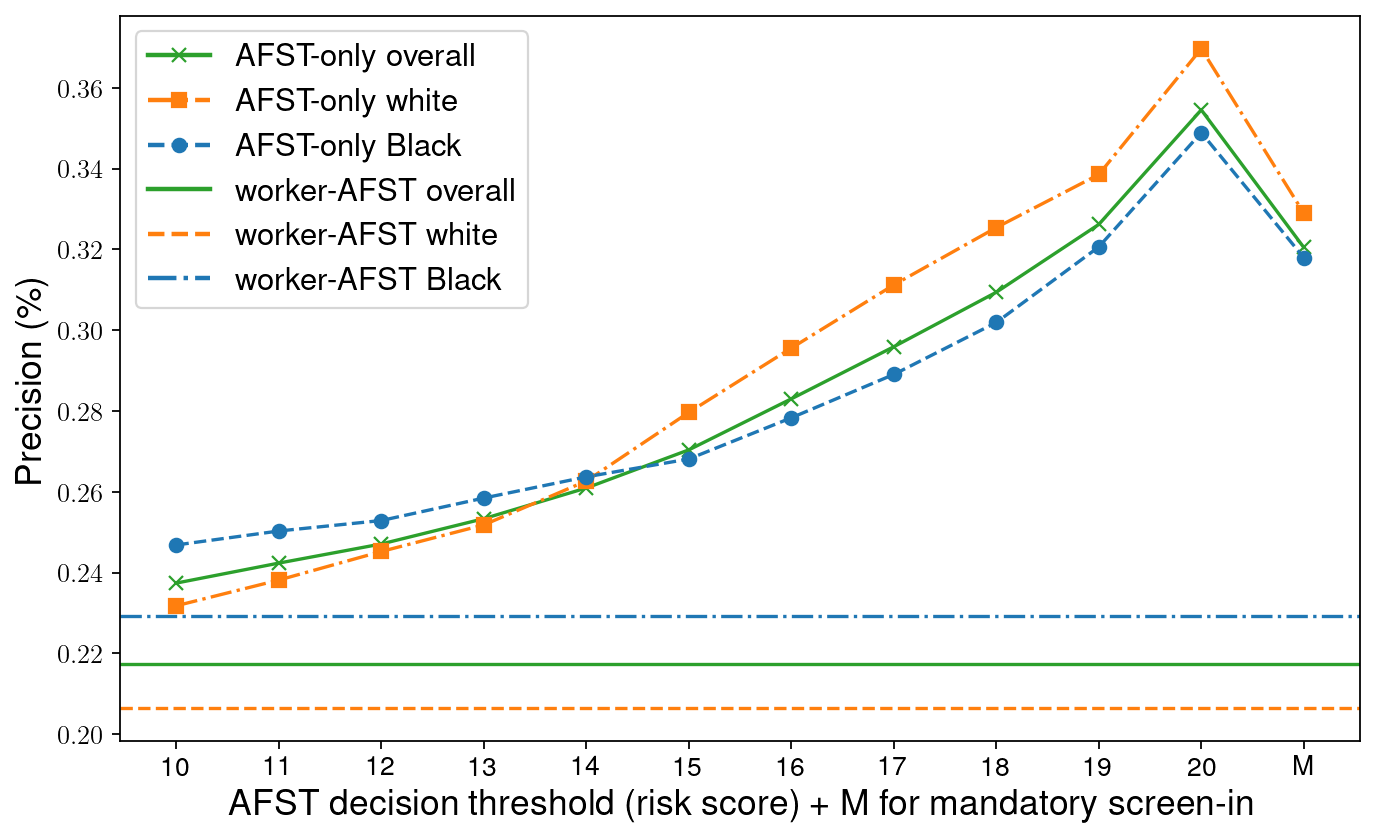} 
        \caption{Precision} \label{fig:per_threshold_precision_analysis1}
    \end{subfigure}
    \hfill
    \begin{subfigure}[t]{0.48\textwidth}
        \centering
        \includegraphics[width=\linewidth]{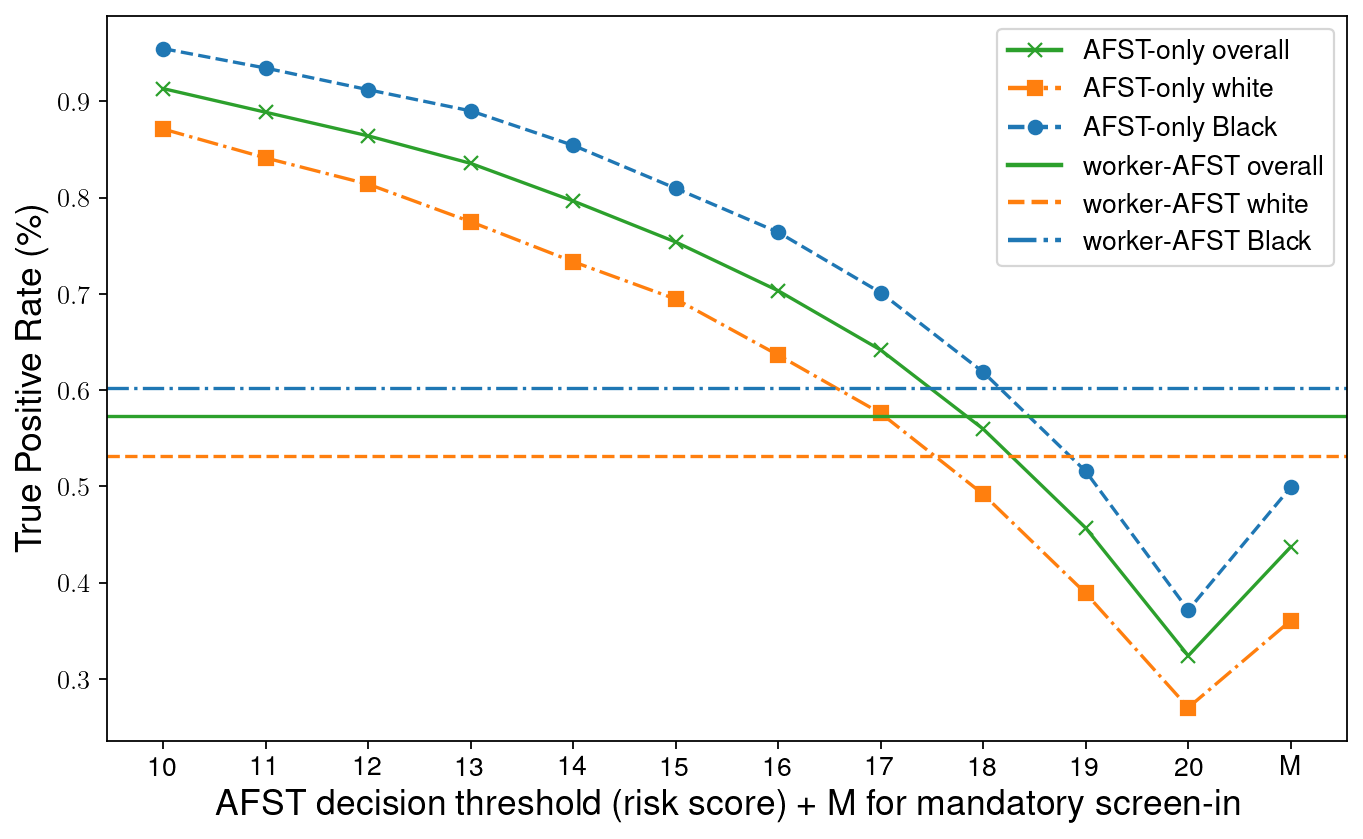} 
        \caption{True positive rate} \label{fig:per_threshold_TPR_analysis1}
    \end{subfigure}
    \begin{subfigure}[t]{0.48\textwidth}
        \centering
        \includegraphics[width=\linewidth]{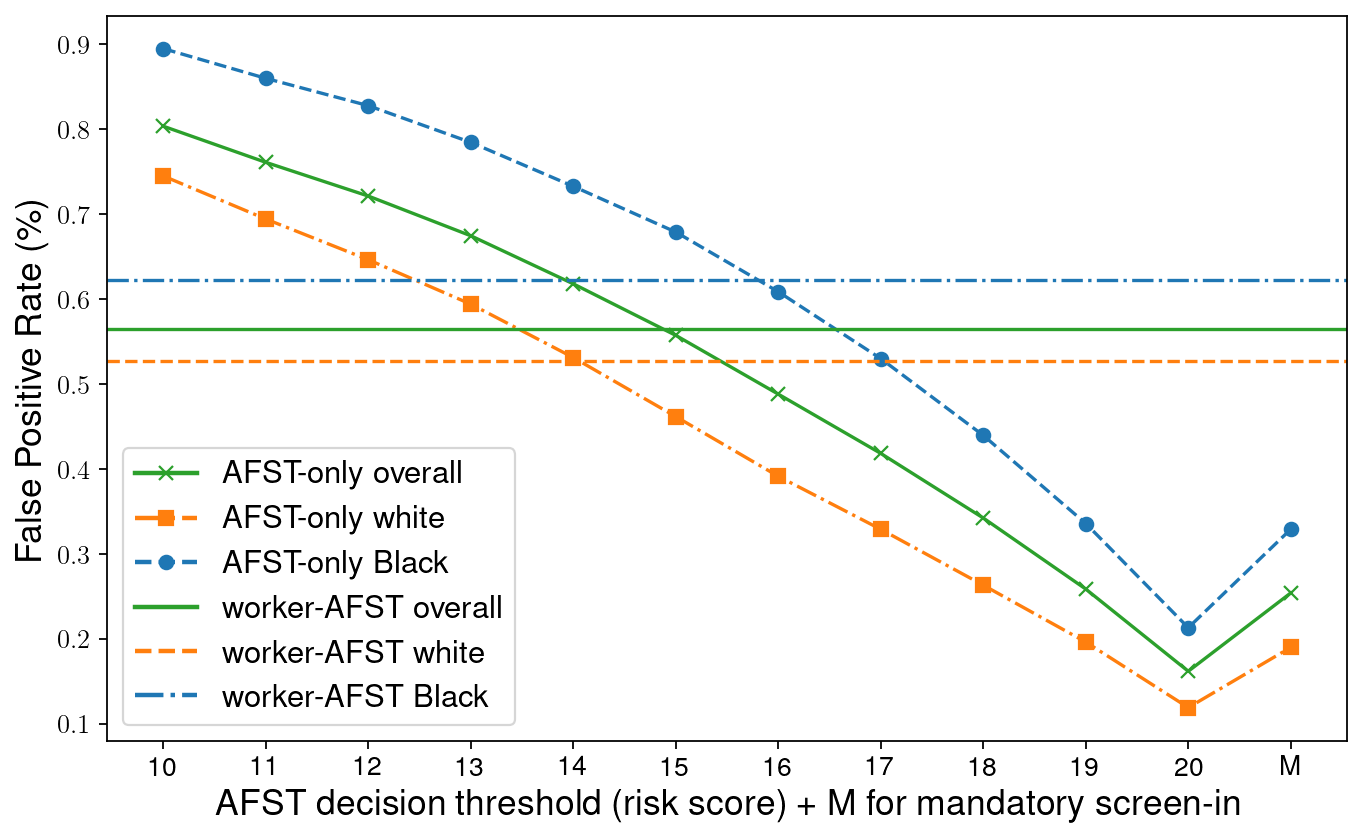} 
        \caption{False positive rate} \label{fig:per_threshold_FPR_analysis1}
    \end{subfigure}
\end{figure*}

\begin{figure*}[h]
    \centering
    \caption{Comparisons of decision outcomes between worker-AFST and AFST-only decisions (Figure 7 in the paper \cite{cheng2022disparities}) based on Analysis 2}
    \label{fig:per_threshold_analysis2}
    \begin{subfigure}[t]{0.48\textwidth}
        \centering
        \includegraphics[width=\linewidth]{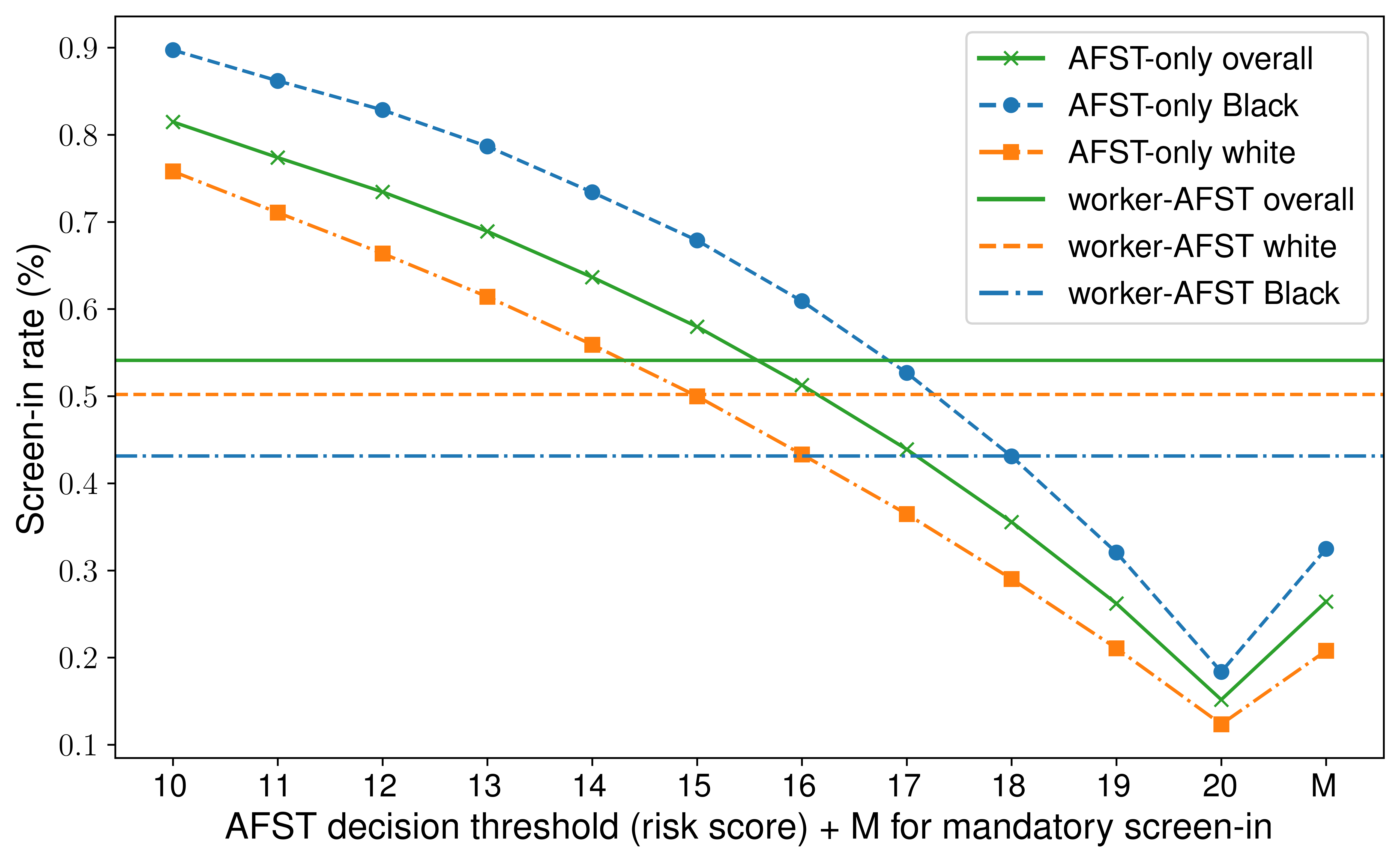} 
        \caption{Screen-in rate} \label{fig:per_threshold_screen_in_rate_analysis2}
    \end{subfigure}
    \hfill
    \begin{subfigure}[t]{0.48\textwidth}
        \centering
        \includegraphics[width=\linewidth]{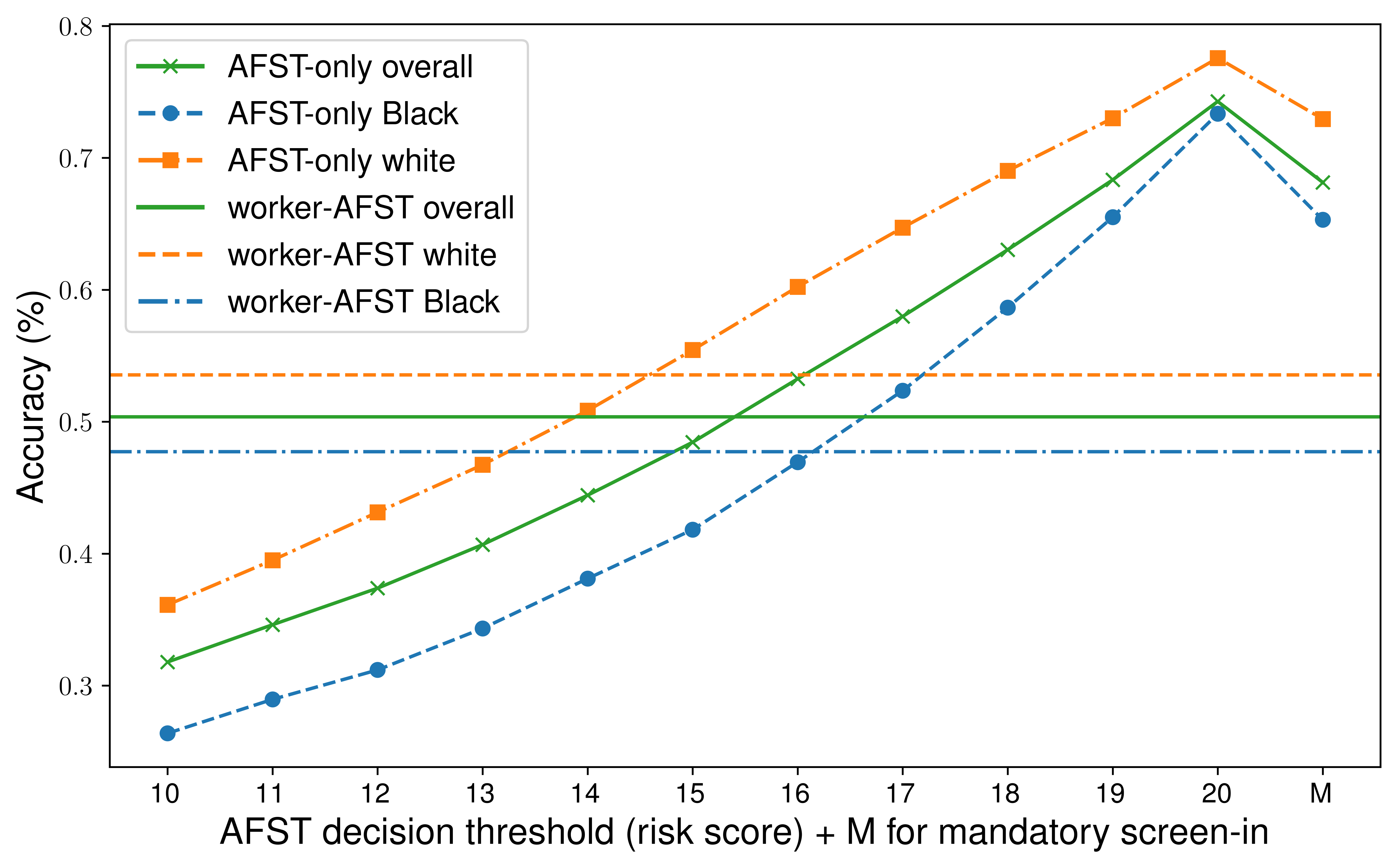} 
        \caption{Accuracy} \label{fig:per_threshold_accuracy_analysis2}
    \end{subfigure}
    \vspace{1cm}
    \begin{subfigure}[t]{0.48\textwidth}
        \centering
        \includegraphics[width=\linewidth]{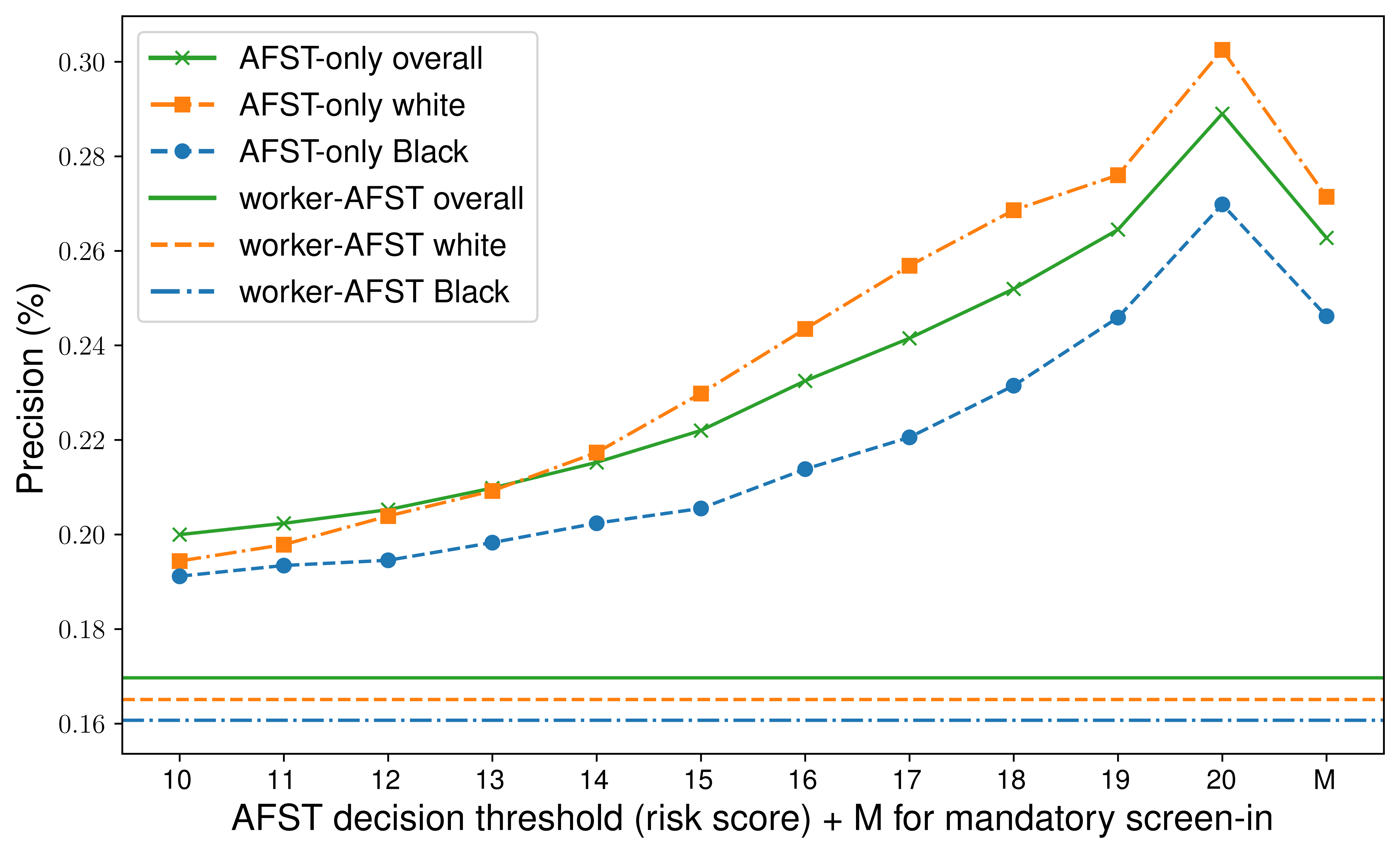} 
        \caption{Precision} \label{fig:per_threshold_precision_analysis2}
    \end{subfigure}
    \hfill
    \begin{subfigure}[t]{0.48\textwidth}
        \centering
        \includegraphics[width=\linewidth]{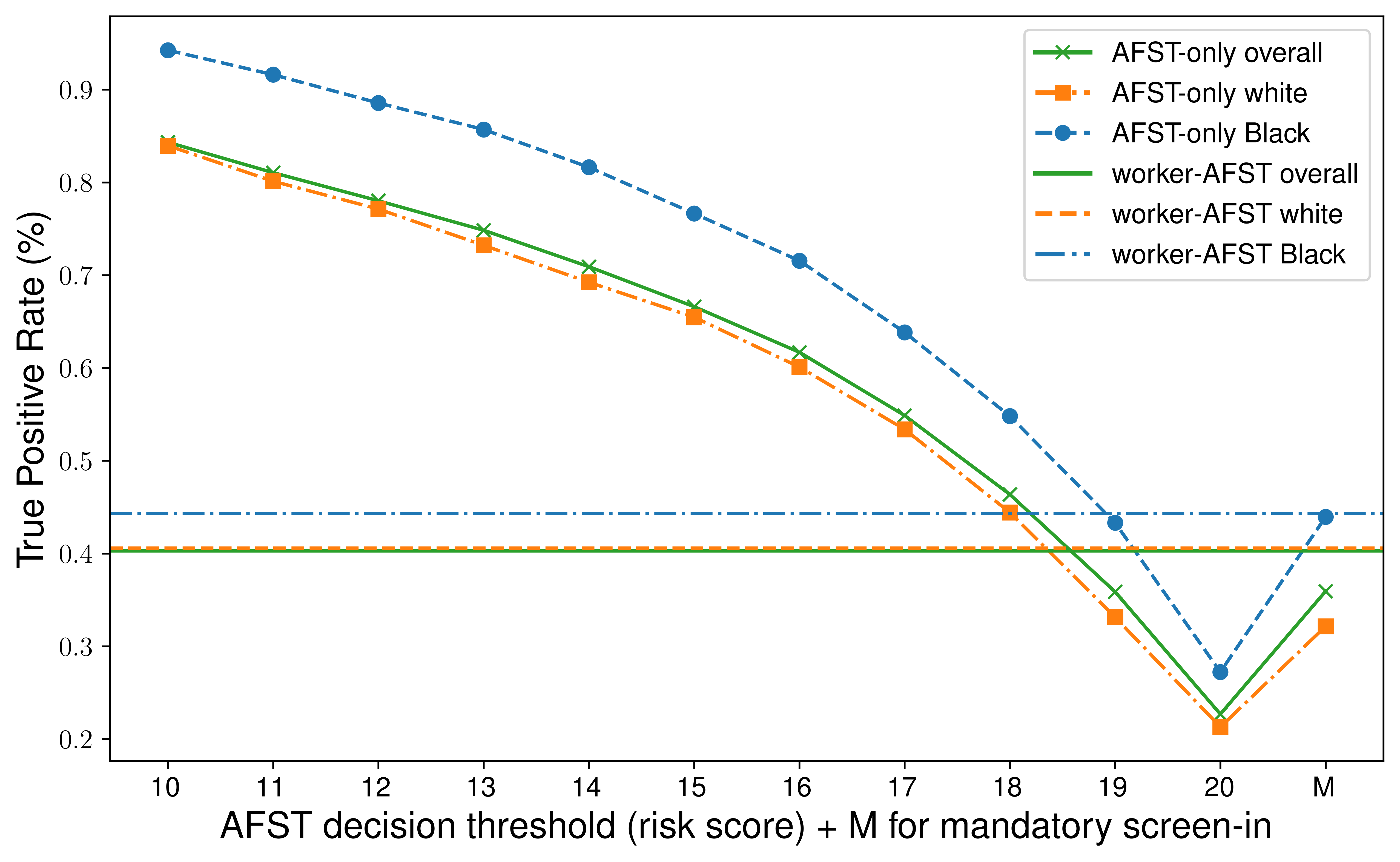} 
        \caption{True positive rate} \label{fig:per_threshold_TPR_analysis2}
    \end{subfigure}
    \begin{subfigure}[t]{0.48\textwidth}
        \centering
        \includegraphics[width=\linewidth]{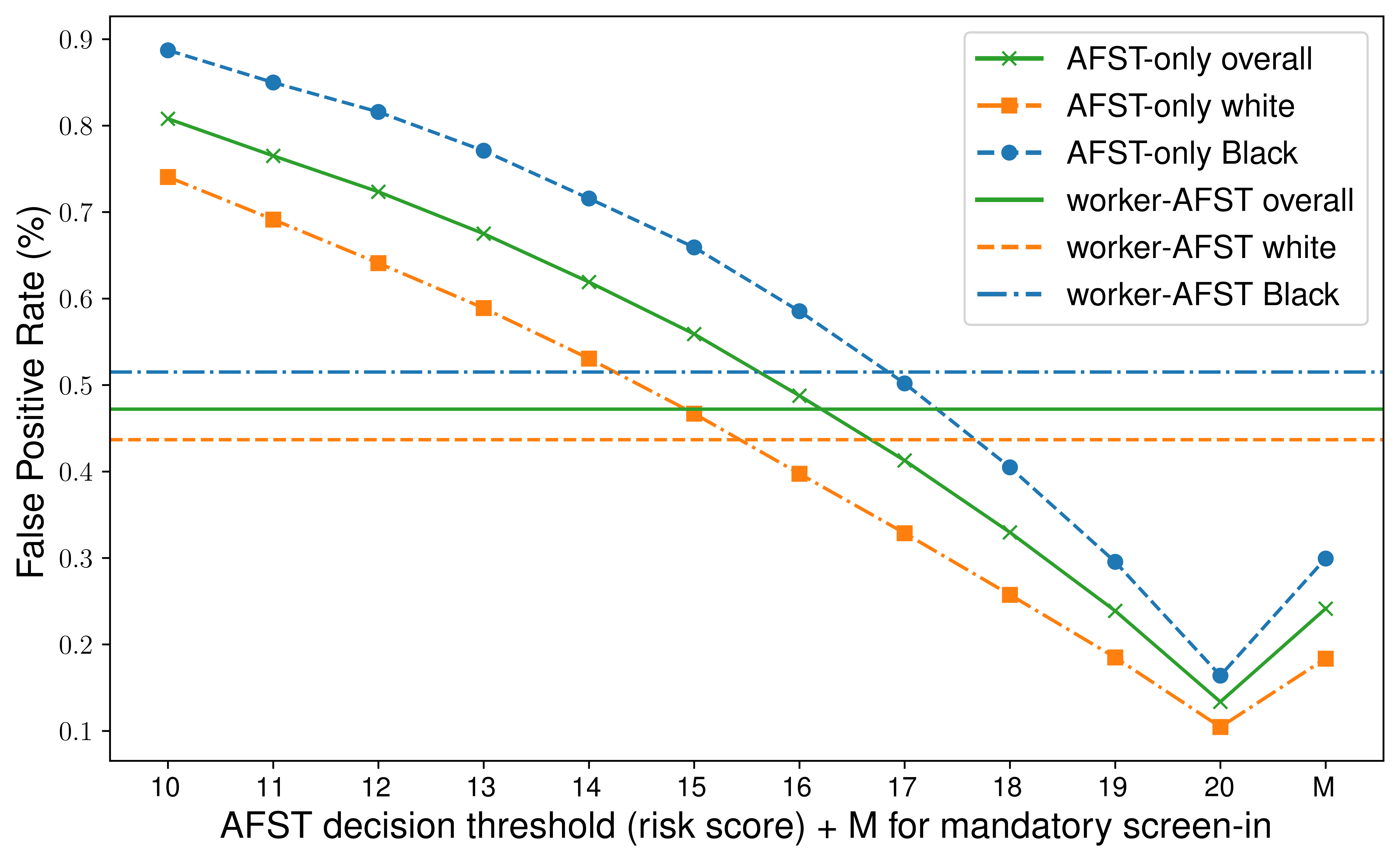} 
        \caption{False positive rate} \label{fig:per_threshold_FPR_analysis2}
    \end{subfigure}
\end{figure*}

\end{document}